\documentclass[aps,pra,onecolumn,superscriptaddress,longbibliography,nofootinbib]{revtex4-2}

\usepackage[T1]{fontenc}
\usepackage[utf8]{inputenc}
\usepackage{amsmath,amssymb,amsthm,mathtools,mathrsfs,bm}
\usepackage{graphicx}
\usepackage{booktabs}
\usepackage{xcolor}
\usepackage{enumitem}
\usepackage{cancel}
\usepackage{tikz} 
\usepackage[colorlinks=true,linkcolor=blue,citecolor=blue,urlcolor=blue]{hyperref}

\newtheorem{definition}{Definition}
\newtheorem{theorem}{Theorem}
\newtheorem{lemma}{Lemma}
\newtheorem{proposition}{Proposition}
\newtheorem{corollary}{Corollary}
\theoremstyle{remark}
\newtheorem{remark}{Remark}

\newcommand{\cO}{\mathcal O}
\newcommand{\cQ}{\mathcal Q}

\newcommand{\cP}{\mathcal P}
\newcommand{\cM}{\mathcal M}
\newcommand{\cH}{\mathcal H}

\newcommand{\bP}{\ensuremath{\mathbf P}}
\newcommand{\bmu}{\ensuremath{\boldsymbol\mu}}
\newcommand{\brho}{\ensuremath{\boldsymbol\rho}}

\newcommand{\btau}{\ensuremath{\boldsymbol\tau}}
\newcommand{\bOmega}{\ensuremath{\boldsymbol\Omega}}
\newcommand{\Tr}{\operatorname{Tr}}
\newcommand{\avgD}{\overline{D}}
\newcommand{\avgS}{\overline{S}}
\newcommand{\KS}{\mathrm{KS}}

\newcommand{\TV}{\mathrm{TV}}
\newcommand{\id}{\mathbb I}
\newcommand{\ket}[1]{|#1\rangle}

\newcommand{\dyad}[1]{|#1\rangle\!\langle#1|}
\newcommand{\abs}[1]{\left|#1\right|}
\newcommand{\set}[1]{\left\{#1\right\}}
\newcommand{\braket}[2]{\langle #1|#2\rangle}

\usetikzlibrary{arrows.meta,positioning,calc}
\newcommand{\bC}{\ensuremath{\bm{C}}}
\begin{document}

\title{Quantum description of reality is epistemically incomplete}

\author{Anubhav Chaturvedi}
\email{anubhav.chaturvedi@ug.edu.pl}
\affiliation{Division of Quantum Optics and Information, Institute of Theoretical Physics and Astrophysics, 
Faculty of Applied Physics and Mathematics, Gda\'nsk University of Technology, 80-233 Gda\'nsk, Poland}

\author{Marcin Paw{\l}owski}
\affiliation{International Centre for Theory of Quantum Technologies (ICTQT), University of Gda\'nsk, 80-308 Gda\'nsk, Poland}

\author{Debashis Saha}
\affiliation{S. N. Bose National Centre for Basic Sciences, Kolkata 700106, India}
\affiliation{Department of Physics, School of Basic Sciences, Indian Institute of Technology Bhubaneswar, Odisha 752050, India}

\begin{abstract}
Whether the operational quantum description is complete can already be asked at the level of preparations: can the empirically accessible properties of a finite preparation set be reproduced exactly by an ontological or hidden-variable description of reality, or must every such completion inevitably contain additional hidden structure that is not operationally accessible? We formalize this question by introducing \emph{epistemic completeness}, a preparation-side notion of classicality requiring that empirical preparation-properties---defined here as optimal one-way communication-task payoffs, and in particular as success probabilities for normalized scoring rules---be preserved exactly by the corresponding not-fine-tuned ontic quantities obtained by conditioning on the ontic state and allowing all response schemes compatible only with positivity and normalization. For the canonical family of set-distinguishability tasks, we prove an exact universal theorem: in every epistemically complete theory, and for every finite preparation set, the average pairwise distinguishability equals the average set-distinguishability. Any nonzero deviation from this equality certifies epistemic incompleteness and quantitatively lower-bounds the excess ontic communication power that every ontic completion must conceal. Because unrestricted classical communication models, and more generally finite-dimensional commuting quantum theories, are epistemically complete, every nonzero deviation from the equality yields a quantum advantage in a one-way communication task under a natural benchmark-task constraint. The same deviations also furnish dimension-independent witnesses of coherence of the preparation set and of measurement incompatibility in any realization of the certifying task values. To characterize the extremal quantum deviations from these equalities, we formulate dimension-independent semidefinite-programming relaxations and dimension-dependent see-saw lower bounds, and show that quantum theory violates the equality in both directions. In particular, we identify the canonical trine and tetrahedral qubit ensembles as the numerically certified dimension-independent maximizers of the positive deviation for the \(n=3\) and \(n=4\) equalities, respectively. We further prove that these positive-direction violations persist for arbitrarily low positive visibility and arbitrarily large leakage short of complete disclosure in the corresponding visibility and leakage families. We also show that the Kochen--Specker \(\psi\)-epistemic model exactly saturates the hidden ontic excess implied by the maximal positive violations for \(n=3\) and \(n=4\), thereby making the witness quantitatively sharp. Finally, our numerical results provide evidence that the maximal positive deviation increases with the number of preparations, thereby motivating the question of the absolute maximal epistemic incompleteness of quantum theory.
\end{abstract}

\maketitle

\section{Introduction}

The question whether the quantum-mechanical description of reality is complete is one of the oldest and deepest questions in quantum foundations \cite{PhysRev.47.777}. In an operational theory, a physical system is described through preparations. The theory further specifies the measurements that can be performed on the system and the rule assigning outcome statistics to each preparation--measurement pair. These statistics determine empirically accessible and falsifiable properties of finite sets of preparations. The completeness question can therefore be posed operationally: do these empirically accessible properties already exhaust the physical content of the operational description, or must any ontological, hidden-variable, or otherwise classical model of reality inevitably contain additional structure that remains inaccessible at the operational level? In the present work we address this question by formalizing it for finite sets of preparations.

This formulation is also motivated by a broader tension at the heart of quantum theory. On the one hand, the theory underlies some of the most powerful sources of advantage in communication, computation, and information processing. On the other hand, there remains no consensus on the precise sense in which the quantum description resists complete classical explanation. These two aspects naturally invite a common question: is the source of quantum advantage connected to the same feature of the formalism that obstructs a complete classical account of physical reality? To ask that question precisely, one needs operationally meaningful notions of classicality. In the realist framework, such notions typically seek classical explanations of empirically accessible operational phenomena in terms of an underlying ontic or hidden-variable description \cite{PhysicsPhysiqueFizika.1.195,10.2307/24902153,PhysRevA.71.052108,Chaturvedi2020quantum,catani2020mathematical}. Bell locality, for example, seeks a realist explanation of non-signalling correlations compatible with local causal structure \cite{PhysicsPhysiqueFizika.1.195,RevModPhys.38.447,Clauser1969}; generalized noncontextuality seeks ontic representations that preserve operational equivalences \cite{PhysRevA.71.052108}; and bounded ontological distinctness seeks preservation of operational distinguishability at the ontic level \cite{Chaturvedi2020quantum}. In each case, an empirically testable operational feature is singled out and required to remain intact in a classical description of reality. The same features, however, also reappear in a second role: when the corresponding classical constraints are violated, one learns that no such classical description can preserve the operational phenomenon exactly, and these very violations can at the same time identify resources that power quantum advantage \cite{RevModPhys.86.419,PhysRevLett.102.010401,PhysRevA.100.022108,Manna2024}. This suggests that if one seeks a more complete classical explanation of the operational theory, one should ask not only whether one selected operational feature can be preserved, but whether the empirically accessible properties of finite sets of preparations can be preserved more generally.

For a finite set of preparations, the operational content of the theory is encoded in the prepare-and-measure statistics that the set can generate against the measurements allowed by the theory. In this work we probe the limits of those statistics through the maximal success probabilities achievable in one-way communication tasks, equivalently the corresponding optimal task values, attainable from them. These quantities are operationally meaningful because they are defined directly from observable prepare-and-measure statistics, and they are empirically falsifiable because any proposed upper bound can be ruled out by an observed performance that exceeds it. We refer to them as \emph{empirical preparation-properties}. 

This level of description is also natural from the perspective of modern prepare-and-measure and semi-device-independent work. Communication-task formulations already play an important role in revealing preparation contextuality, certifying classicality directly from observable prepare-and-measure behaviors, deriving structural criteria for arbitrary prepare-and-measure scenarios, and self-testing communicated states \cite{PhysRevLett.119.220402,Ambainis2019,PRXQuantum.2.030311,GittonWoods2022,Saha_2019,SelfTestingPM2023}. On the semi-device-independent side, many physically motivated assumptions on the communicated system are not imposed microscopically, but are instead formulated, compared, or relaxed through information-like quantities associated with the same set of preparations \cite{VanHimbeeck2017,Tavakoli2020informationally,Tavakoli2022informationally,PauwelsPironioTavakoli2025}. In particular, recent work has clarified that one-shot accessible information provides a common relaxation of a broad class of natural physical assumptions, and that the corresponding optimal discrimination probability is an especially central operational quantity in this setting \cite{Chaturvedi2020quantum,Tavakoli2020informationally,PauwelsPironioTavakoli2025}. Since distinguishability and related guessing probabilities are themselves communication-task performance measures of the communicated preparations, the present work asks what exact constraints complete classical explanation imposes when one considers the full family of empirical communication-task quantities associated with a finite preparation set.

The corresponding comparison is between two levels of description. At the operational level, a finite preparation set is characterized by empirical communication-task quantities obtained by optimizing over the measurements allowed by the operational theory. At the ontic or hidden-variable level, conditioned on access to the ontic state, one may ask for the corresponding optimal task values when all response schemes compatible only with positivity and normalization are allowed. These are the \emph{not}-fine-tuned ontic quantities relevant for the present completeness question. In this language, complete classical explanation means exact preservation of empirical preparation-properties at the ontic level. Conversely, if the unrestricted ontic quantity exceeds the operational one, then the ontic completion contains additional communication power that is not operationally accessible, and any recovery of the operational description necessarily requires fine-tuning in the form of a restriction to a strict subset of the valid ontic decoders.

Motivated by this viewpoint, we introduce a notion of classicality called \emph{epistemic completeness}. In the present setting, an operational theory is epistemically complete if, for every finite set of preparations, each empirically accessible preparation-property is reproduced exactly by the corresponding not-fine-tuned ontic quantity of some underlying epistemic states, namely probability distributions over underlying states of reality. The notion is deliberately broad: it does not protect one singled-out operational witness or one distinguished relation among preparations, but asks whether the empirically accessible communication behavior of finite preparation sets is already completely classically explainable.

At the same time, the central theorem of the paper already follows from one canonical family inside this broader class, namely the family of set-distinguishability tasks. In these tasks, Bob must output an $m$-element set containing the true preparation label. The endpoint cases recover two familiar operational quantities: for $m=1$, one has ordinary distinguishability \cite{Chaturvedi2020quantum}, while for $m=n-1$, one has antidistinguishability \cite{j54v-ng1h}. Intermediate values of $m$ interpolate between exact identification and exact exclusion, and therefore organize, in a canonical way, how the same preparation set performs across a family of communication tasks of increasing coarseness.

From this family one extracts two natural averaged quantities. The first is the average pairwise distinguishability, which records, on average, how well the preparations can be discriminated two at a time. The second is the average set-distinguishability, which averages performance across all nontrivial output-set sizes and therefore captures the broader communication behavior of the same preparation set. The central theorem of the manuscript is that these two quantities must coincide in every epistemically complete theory. More precisely, for every finite preparation set, the average pairwise distinguishability equals the average set-distinguishability.

This exact equality is the main witness introduced in the paper. Its force is universal and structural. It holds for every finite preparation set in every epistemically complete theory, and it holds irrespective of the particular underlying epistemic states. The proof is exact and elementary at its core: once one passes to unrestricted ontic decoding, the equality follows from a simple identity for ordered lists of real numbers. The two averaged quantities are therefore not independent constraints imposed from outside, but two operational shadows of the same unrestricted ontic data. In this sense, the equality is a preparation-side signature of complete classical explanation.

The witness also has immediate quantitative, operational, and structural meaning. Any nonzero deviation certifies epistemic incompleteness: its magnitude lower-bounds the amount of hidden ontic communication power that every ontic or hidden-variable completion must conceal, while its sign identifies where that hidden excess must appear. In this sense, the witness is an accounting principle for hidden ontic task-power. 

At the same time, because unrestricted classical communication models are epistemically complete, every nonzero deviation yields a quantum advantage in a benchmark-constrained one-way communication task, in a manner consonant with related contextuality-based communication and distinguishability-based quantum advantages \cite{PhysRevA.100.022108,PhysRevA.109.032212,Manna2024,Ray2025}. If the deviation is positive, then under the benchmark constraint fixing the corresponding average set-distinguishability, the given preparations attain an average pairwise distinguishability that no classical model can match; if the deviation is negative, then under the benchmark constraint fixing the average pairwise distinguishability, the same preparations attain an average set-distinguishability beyond the classical limit. More generally, because every finite-dimensional commuting quantum theory is epistemically complete, any nonzero deviation is also a dimension-independent witness that the given set of preparations cannot be jointly realized within any finite-dimensional commuting quantum theory, and therefore certifies coherence of the preparation set. Likewise, in any concrete realization of the certifying task values, a nonzero deviation rules out a jointly measurable certifying family of measurements, since such a realization would again reduce to an effectively classical or commuting quantum theory and would therefore satisfy the equality. Thus a single nonzero deviation simultaneously quantifies epistemic incompleteness, implies benchmark-constrained quantum communication advantage, certifies coherence of the preparations, and witnesses measurement incompatibility in any realization that attains it.

We then show that quantum theory violates the equality in both directions. To characterize the extremal deviations, we formulate dimension-independent semidefinite-programming relaxations together with dimension-dependent see-saw lower bounds, following the moment-matrix and hierarchy methods developed for contextuality and informationally restricted prepare-and-measure scenarios \cite{Chaturvedi2021characterising,PRXQuantum.2.020334,Tavakoli2022informationally,Hazra2026efficient}. On the positive side, these identify the canonical trine and tetrahedral qubit ensembles as the exact dimension-independent maximizers of the deviation for the \(n=3\) and \(n=4\) equalities, respectively (up to machine precision). The corresponding positive boundary is then described analytically by two explicit continuations of these canonical qubit points: visibility families obtained by depolarizing the canonical ensembles, and leakage families obtained by adjoining an orthogonal branch that reveals the preparation label. On the negative side, the same methods numerically recover opposite-sign extremal deviations, already realized by qutrit ensembles for \(n=3\) and by higher-dimensional ensembles for \(n=4\). As a consequence of the explicit visibility and leakage families, we further prove analytically that the \(n=3\) and \(n=4\) positive violations persist for arbitrarily low positive visibility and arbitrarily large leakage short of complete disclosure. Thus, the operational quantum description is epistemically incomplete already on simple and highly symmetric qubit fragments, and that incompleteness remains robust deep into noisy and high-leakage regimes. 

Using the numerical tools developed above, we also find evidence that the maximal positive deviation increases with the number of preparations in the cases studied. This leads naturally to the question of the absolute maximal epistemic incompleteness of quantum theory, namely the largest possible equality-witness deviation over all finite preparation sets and all \(n\).

The Kochen--Specker \(\psi\)-epistemic model occupies a distinguished place in the present work. Historically, it is the canonical \(\psi\)-epistemic hidden-variable model for pure qubit states and projective measurements \cite{10.2307/24902153,Quanta22}. For the canonical trine and tetrahedral ensembles, we show that it exactly saturates the hidden ontic excess implied by the equality violation. Thus, for these paradigmatic qubit examples, the equality witness is not merely a no-go statement against complete classical explanation: it is quantitatively sharp. In this precise sense, the Kochen--Specker model identifies the exact amount of additional ontic communication power that ontological completion must expose in these extremal cases.

The present framework also clarifies the place of earlier notions of classicality by showing that epistemic completeness provides a common preparation-side umbrella under which they can be organized. Since ordinary distinguishability is one empirical preparation-property among many, epistemic completeness immediately subsumes bounded ontological distinctness on preparations \cite{Chaturvedi2020quantum}. In this sense, it is the natural generalization of preparation-side distinguishability-based classicality from one selected operational quantity to the full family of empirical communication-task quantities of a finite preparation set. This already links it directly to preparation noncontextuality and, under the standard additional structural assumptions familiar from the ontological-model literature, places it within the broader hierarchy connecting preparation-side assumptions to maximally \(\psi\)-epistemic structure, Kochen--Specker noncontextuality, and, in the wider realist setting, Bell local causality \cite{PhysRevLett.110.120401,PhysRevLett.112.250403,Quanta22}. Antidistinguishability, which appears here as the endpoint \(m=n-1\) of the same task family, likewise has a well-established role in contextuality and noncontextuality inequalities \cite{LeiferDuarte2020,Srikumar2026,Johnston2025,j54v-ng1h}. The family of set-distinguishability tasks studied here therefore does not merely supply one more witness, but unifies within a single preparation-side framework the operational ingredients from which several earlier notions of classicality arise.

The same operational language also leads naturally to a complementary framework for benchmark-constrained nonclassicality. Here the question is one of exact preservation under complete classical explanation. But the very same empirical communication-task quantities can also be used operationally, by placing benchmark bounds on some tasks and witnessing nonclassicality through performance in others. The present framework therefore points directly to a broader constrained-communication perspective in which communication tasks play a dual role: they function simultaneously as witnesses and as constraints.

The paper is organized as follows. Section II introduces prepare-and-measure theories and empirical preparation-properties. Section III formulates the ontic comparison, defines not-fine-tuned ontic quantities and epistemic completeness, and discusses two epistemically complete comparison classes: unrestricted classical communication and finite-dimensional commuting quantum theories. Section IV introduces the family of set-distinguishability tasks and the two averaged quantities extracted from it. Section V proves the exact equality theorem and derives the quantitative lower bound on hidden ontic excess implied by any violation. Section VI develops the general implications of a nonzero deviation, including benchmark-constrained quantum communication advantage, coherence witnesses for preparation sets, and measurement-incompatibility witnesses for any realizing implementation. Section VII presents the quantum violations, including the exact trine and tetrahedral examples, the geometric picture in the \((\avgS_n,\avgD_n)\) plane displaying both signs of deviation for \(n=3\), the numerical extremality of the canonical qubit ensembles for the \(n=3\) and \(n=4\) equalities, and the negative-direction extremum for \(n=3\). Section VIII studies the explicit visibility and leakage families and proves the corresponding robustness theorems. Section IX shows that the Kochen--Specker \(\psi\)-epistemic model exactly saturates the hidden ontic excess for the canonical qubit ensembles, thereby making the equality witness quantitatively sharp. Section X formulates the semidefinite-programming relaxations and see-saw lower bounds for the maximal quantum deviation and discusses the evidence that the maximal positive deviation increases with \(n\). Section XI places epistemic completeness within the broader hierarchy of notions of classicality and explains its relation to the complementary benchmark-constrained framework in which communication tasks function both as witnesses and as constraints. We conclude in Sec.~XII.

\section{Prepare-and-measure theories and empirical preparation-properties}

\subsection{Prepare-and-measure experiments}

We begin with the simplest operational primitive needed in this work: a prepare-and-measure experiment. An operational theory \(\cO\) specifies a set of preparations \(\cP_{\cO}\), a set of measurements \(\cM_{\cO}\), and for every preparation \(P\in\cP_{\cO}\) and measurement \(M\in\cM_{\cO}\), a conditional probability distribution
\begin{equation}
p(k|P,M)
\end{equation}
for the measurement outcomes \(k\).

This elementary framework already suffices for the present completeness question. The reason is that the issue studied here is entirely preparation-side: for a given finite set of preparations, what empirically accessible properties are assigned to it by the operational theory, and what additional communication power would become available if an ontic or hidden-variable completion made the underlying state accessible? No transformations, multipartite structure, or further operational ingredients are needed for that comparison.

In quantum theory, a preparation is represented by a density operator \(\rho\) on a finite-dimensional Hilbert space \(\cH\), a measurement is represented by a POVM \(M=\set{M_k}_k\) with \(M_k\succeq0\) and \(\sum_k M_k=\id\), and the predicted statistics are given by the Born rule
\begin{equation}
p(k|\rho,M)=\Tr(\rho M_k).
\end{equation}

A second operational model that will play a central comparative role is unrestricted classical communication. In that model, a preparation is an encoding scheme \(e_x(\omega)\) over a finite classical message alphabet, and a measurement is a decoding scheme \(d_M(k|\omega)\). The induced operational statistics take the form
\begin{equation}
p(k|x,M)=\sum_\omega e_x(\omega)\,d_M(k|\omega).
\end{equation}
As we show below, this class of models is epistemically complete and therefore provides the natural classical benchmark for the framework developed here. With this operational setting fixed, the next step is to specify which preparation-side quantities we will treat as the empirically accessible properties whose exact preservation is at stake.

\subsection{Communication tasks as empirical preparation-properties}

Let \(\bP=(P_1,\dots,P_n)\) be an ordered set of \(n\) preparations. Associated with \(\bP\) we consider task-values of the form
\begin{equation}
\label{eq:generic-task}
S_n^{(\cO)}(\bP)=\max_{M\in\cM_{\cO}}\sum_{x,k} c_k^x\, p(k|P_x,M),
\end{equation}
where the coefficients \(c_k^x\in\mathbb R\) specify a one-way communication task. For arbitrary real coefficients \(c_k^x\), Eq.~\eqref{eq:generic-task} defines a linear task-payoff functional. In the special case where the coefficients specify a normalized scoring rule, it is an optimal success probability.

This quantity should be read operationally. The coefficients encode the scoring rule of the task. Alice chooses a label \(x\) and sends the preparation \(P_x\). Bob performs a measurement and outputs \(k\). For a fixed measurement, the achieved score is \(\sum_{x,k} c_k^x p(k|P_x,M)\). Maximizing over all measurements allowed by the theory removes dependence on any one particular decoder and leaves a quantity determined only by the preparation set itself.

In this way, Eq.~\eqref{eq:generic-task} defines an \emph{empirical preparation-property}. It is a preparation-property because, after the maximization, it depends only on the set \(\bP\), not on a specific measurement choice. It is empirical because it is defined directly from observable prepare-and-measure statistics. This point is crucial for the present framework: if such quantities are to function as physical prerequisites, witnesses, or benchmark constraints, they must be operationally falsifiable. The task-values considered here have precisely that status, since any claimed upper bound on \(S_n^{(\cO)}(\bP)\) can in principle be tested and falsified by observing a larger task performance.

This class is deliberately broad. It contains minimum-error discrimination, exclusion tasks, set-distinguishability tasks, parity-type tasks, and many other one-way communication tasks. It therefore provides a natural operational language in which to formulate the completeness question for preparations.

Although the framework applies to a broad class of empirical preparation-properties, the central theorem of the paper already follows from one canonical subfamily. We therefore specialize next to subset-identification, equivalently set-distinguishability, tasks, which interpolate between ordinary distinguishability and antidistinguishability.
\subsection{Subset-identification tasks}

Among all tasks of the form \eqref{eq:generic-task}, a distinguished role will be played by subset-identification tasks. Fix an integer $m\in\{1,\dots,n-1\}$. In the $(n,m)$ subset-identification task, Bob must output an $m$-element subset of $[n]=\{1,\dots,n\}$ that contains the true label $x$. The corresponding optimal success probability is
\begin{equation}
\label{eq:Dnm-op}
D_{n,m}^{(\cO)}(\bP)
=
\frac1n
\max_M
\sum_{\substack{I\subset[n]\\ |I|=m}}
\ \sum_{x\in I}
p(k=I|P_x,M).
\end{equation}

This family interpolates naturally between several well-known tasks. When $m=1$, Bob must identify the label exactly, and we recover ordinary distinguishability:
\begin{equation}
D_{n,1}^{(\cO)}(\bP)
=
\frac1n \max_M \sum_{x=1}^n p(k=x|P_x,M).
\end{equation}
At the other end, when $m=n-1$, Bob must output a subset omitting one label, so the task becomes one of exclusion or antidistinguishability. Thus the same family contains both exact identification and exact exclusion as special cases.

Intermediate values of $m$ are equally natural. They ask how well a communicated preparation enables Bob to narrow the label down to a subset of prescribed size. From the present perspective, this family is especially important because it provides a canonical hierarchy of communication tasks built from the same preparation set.

\subsection{Average pairwise distinguishability and average set distinguishability}

We now define the two aggregated preparation-properties that form the central equality witness of the paper.

The \emph{average pairwise distinguishability} is
\begin{equation}
\label{eq:avgD-def}
\avgD_n^{(\cO)}(\bP)
:=
\frac{1}{\binom{n}{2}}
\sum_{1\le i<j\le n}
D_{2,1}^{(\cO)}(P_i,P_j).
\end{equation}
This quantity averages the best binary discrimination success over all unordered pairs of preparations. It tells us, on average, how well the preparations can be separated two at a time.

The \emph{average set distinguishability} is
\begin{equation}
\label{eq:avgS-def}
\avgS_n^{(\cO)}(\bP)
:=
\frac{1}{n-1}
\sum_{m=1}^{n-1} D_{n,m}^{(\cO)}(\bP).
\end{equation}
This quantity averages the best success probability across the whole subset-identification hierarchy. It therefore captures the broader task-profile of the entire preparation set.

The equality witness compares these two ways of aggregating preparation-side communication power:
\begin{equation}
\avgD_n^{(\cO)}(\bP)\quad\text{versus}\quad \avgS_n^{(\cO)}(\bP).
\end{equation}
The remarkable fact proved below is that for epistemically complete theories they must coincide exactly for every preparation set. Any deviation is therefore a direct witness of incompleteness.

With the empirical preparation-properties now in place, the next step is to lift them to the hidden-variable level and to make precise what it means for their preservation to fail only through fine-tuning. We therefore turn next to ontic extensions, ontological models, and the definition of epistemic completeness itself.

\section{Ontic extensions, fine-tuning, and epistemic completeness}

\subsection{Ontic extensions}

To ask whether the operational preparation description is complete, one must first say what kind of hidden-level extension is being considered. The weakest notion needed for the present work is that of an ontic extension \cite{catani2020mathematical}.

\begin{definition}[Ontic extension]
An ontic extension $Q$ of a prepare-and-measure fragment of an operational theory $\cO$ specifies, for every preparation $P$, measurement $M$, and outcome $k$, a joint conditional distribution
\begin{equation}
q(k,\lambda|P,M)
\end{equation}
over the operational outcome $k$ and an ontic variable $\lambda\in\Lambda$, such that
\begin{equation}
p(k|P,M)=\int_\Lambda d\lambda q(k,\lambda|P,M).
\end{equation}
\end{definition}

This notion is weaker than a full ontological model. It does not yet insist that $\lambda$ mediates all correlations between preparation and measurement devices, nor that the ontic state be independent of the choice of measurement. It simply says that the operational description admits an extension by a hidden variable.

This weaker notion is useful because the key issue of the paper is already visible here: once $\lambda$ is available, what additional task-power becomes logically accessible?

\subsection{Ontological models as a special case}

A standard ontological model is obtained by supplementing an ontic extension with two familiar structural assumptions.

First, one requires \emph{$\lambda$-mediation}: once the ontic state is specified, the operational outcome should depend on the preparation only through $\lambda$,
\begin{equation}
\label{eq:lambda-med}
q(k|P,M,\lambda)=\xi_M(k|\lambda).
\end{equation}

Second, one requires \emph{measurement independence}: the ontic state produced by a preparation does not depend on which measurement will later be chosen,
\begin{equation}
\label{eq:meas-ind}
q(\lambda|P,M)=\mu_P(\lambda).
\end{equation}

With these assumptions the joint distribution factorizes as
\begin{equation}
q(k,\lambda|P,M)=\mu_P(\lambda)\,\xi_M(k|\lambda),
\end{equation}
which is the standard ontological-model form \cite{PhysRevA.71.052108,Quanta22}.

\subsection{Not-fine-tuned ontic task-values}

We now define the ontic counterpart of an empirical preparation-property. The guiding idea is this: if the hidden state of reality $\lambda$ were accessible, then the decoder should be allowed to use \emph{every} response scheme constrained only by positivity and normalization, not merely the smaller subset realized operationally by the theory.

\begin{definition}[Not-fine-tuned ontic task-value]
Given an ontic state space $\Lambda$ and an associated family of epistemic states $\bmu=(\mu_1,\dots,\mu_n)$, the not-fine-tuned ontic task-value associated with the coefficients $\set{c_k^x}$ is
\begin{equation}
\label{eq:ontic-task-def}
S_n^{(\Lambda)}(\bmu)
:=
\max_{\xi\in\Xi_\Lambda}
\sum_{x,k}
c_k^x
\int_\Lambda d\lambda, \mu_x(\lambda)\,\xi(k|\lambda),
\end{equation}
where $\Xi_\Lambda$ denotes the set of all response schemes $\xi(k|\lambda)$ satisfying positivity and normalization.
\end{definition}

Because the allowed response schemes are constrained only by positivity and completeness, the maximization in \eqref{eq:ontic-task-def} can be solved pointwise.

\begin{lemma}
\label{lem:ontic-solved}
For every coefficient family $\set{c_k^x}$ and every epistemic ensemble $\bmu$,
\begin{equation}
\label{eq:ontic-task-solved}
S_n^{(\Lambda)}(\bmu)
=
\int_\Lambda d\lambda\,
\max_k
\sum_x c_k^x\,\mu_x(\lambda).
\end{equation}
\end{lemma}

\begin{proof}
For fixed $\lambda$, the optimization problem in \eqref{eq:ontic-task-def} is linear in the response probabilities $\xi(k|\lambda)$ over the simplex of all normalized outcome distributions. Therefore the optimum is attained at an extreme point of that simplex, namely at a deterministic response rule that outputs one $k$ with probability $1$. The optimal deterministic choice is the one maximizing $\sum_x c_k^x \mu_x(\lambda)$. Performing this choice independently for each $\lambda$ yields \eqref{eq:ontic-task-solved}.
\end{proof}

Equation \eqref{eq:ontic-task-solved} is the basic no-fine-tuning benchmark of the paper. It is what the task-value would be if the hidden state could be used without any further restriction.

\subsection{What fine-tuning means here}

The language of fine-tuning is often used in quantum foundations, but it is important to say very clearly what it means in the present context.

At the operational level, the decoder is restricted to the measurement devices available in the operational theory. Thus only a certain subset of response schemes is realizable. At the ontic level, once $\lambda$ is assumed available, there is no logical reason to restrict the decoder beyond positivity and normalization. The not-fine-tuned task-value therefore uses the full set $\Xi_\Lambda$.

If, for some preparation set and some task, the operational value is strictly smaller than the not-fine-tuned ontic value, then the ontic completion contains additional task-power that is not accessible operationally. In the present paper we call this a \emph{fine-tuning} of the completion: the operational theory can only be recovered if the realizable decoders form a strict subset of all decoders that would otherwise be compatible with access to the hidden state. This use of the term is thus very concrete. Fine-tuning means hidden task-power must be concealed. No fine-tuning means empirical preparation-properties are preserved exactly at the ontic level.

This observation leads directly to the central notion introduced in this work. The requirement that empirical preparation-properties be preserved exactly in the above not-fine-tuned sense is what we formalize as epistemic completeness.

\begin{figure}[t]
\centering
\begin{tikzpicture}[
    font=\small,
    title/.style={font=\small\bfseries, align=center},
    frame/.style={draw=black!55, rounded corners=3pt, line width=0.8pt, inner sep=5pt, align=center},
    topbox/.style={frame, fill=blue!4, draw=blue!50!black, text width=3.35cm, minimum height=1.25cm},
    descbox/.style={frame, fill=green!4, draw=green!45!black, text width=3.35cm, minimum height=0.88cm},
    mathbox/.style={frame, fill=black!2, draw=black!45, text width=3.35cm, minimum height=1.62cm},
    keybox/.style={draw=black!45, rounded corners=3pt, fill=black!2, inner sep=6pt, align=center, text width=5.6cm, font=\scriptsize}
]

\node[title] at (-4.25,2.7) {Preparation\\noncontextuality};
\node[title] at (0,2.7) {Bounded ontological\\distinctness};
\node[title] at (4.25,2.7) {Epistemic\\completeness};

\node[topbox] at (-4.25,1.15)
{$P_0 \simeq P_1$\\[0.8mm]
$\Longrightarrow\ D_{2,1}^{(\cO)}(P_0,P_1)=\tfrac12$};

\node[descbox] at (-4.25,-0.35)
{{\color{green!45!black}\bfseries preserve indistinguishability}};

\node[mathbox] at (-4.25,-1.95)
{$\mu_0(\lambda)=\mu_1(\lambda)\ \forall\lambda$\\[0.9mm]
$\Longrightarrow\ D_{2,1}^{(\Lambda)}(\mu_0,\mu_1)=\tfrac12$};

\node[topbox] at (0,1.15)
{$\bP=(P_1,\dots,P_n)$};

\node[descbox] at (0,-0.35)
{{\color{green!45!black}\bfseries preserve distinguishability}};

\node[mathbox] at (0,-1.95)
{$D_{n,1}^{(\cO)}(\bP)=D_{n,1}^{(\Lambda)}(\bmu)$};

\node[topbox] at (4.25,1.15)
{$\bP=(P_1,\dots,P_n)$};

\node[descbox] at (4.25,-0.35)
{{\color{green!45!black}\bfseries preserve every empirical preparation-property}};

\node[mathbox] at (4.25,-1.95)
{$S_n^{(\cO)}(\bP)=S_n^{(\Lambda)}(\bmu)$};

\node[keybox] at (-3.0,-4.2)
{$\begin{aligned}
D_{n,m}^{(\cO)}(\bP)
&=\frac1n\max_M\sum_{\substack{I\subset[n]\\ |I|=m}}\ \sum_{x\in I} p(k=I\mid P_x,M),\\[1.2mm]
D_{n,m}^{(\Lambda)}(\bmu)
&=\frac1n\int_\Lambda d\lambda\,\max_{\substack{I\subset[n]\\ |I|=m}}\ \sum_{i\in I}\mu_i(\lambda)
\end{aligned}$};

\node[keybox] at (3.0,-4.2)
{$\begin{aligned}
S_n^{(\cO)}(\bP)
&=\max_{M\in\cM_{\cO}}\sum_{x,k} c_k^x\,p(k\mid P_x,M),\\[1.2mm]
S_n^{(\Lambda)}(\bmu)
&=\int_\Lambda d\lambda\,\max_k\sum_x c_k^x\,\mu_x(\lambda)
\end{aligned}$};

\end{tikzpicture}
\caption{Three preparation-side preservation principles arranged in increasing strength. Left: preparation noncontextuality preserves operational indistinguishability. Here $P_0\simeq P_1$ denotes operational equivalence. If $P_0\simeq P_1$, then the operational two-state distinguishability is trivial, $D_{2,1}^{(\cO)}(P_0,P_1)=\tfrac12$, and preparation noncontextuality sends this to identical epistemic states $\mu_0(\lambda)=\mu_1(\lambda)$ for all $\lambda$, which in turn fixes the ontic distinguishability at $D_{2,1}^{(\Lambda)}(\mu_0,\mu_1)=\tfrac12$. Middle: bounded ontological distinctness preserves distinguishability itself for an arbitrary finite preparation set $\bP$, by requiring exact agreement between the operational and ontic quantities $D_{n,1}^{(\cO)}(\bP)$ and $D_{n,1}^{(\Lambda)}(\bmu)$; see Proposition~\textup{\ref{prop:ec-implies-bod}}. Right: epistemic completeness strengthens this further by requiring exact preservation of every empirical preparation-property $S_n^{(\cO)}(\bP)$ of the form introduced in Eq.~\textup{\eqref{eq:generic-task}}, through equality with its unrestricted ontic counterpart $S_n^{(\Lambda)}(\bmu)$. The two formula panels at the bottom are written at matched levels. The left panel records the canonical set-distinguishability quantities in their operational and pointwise-solved ontic forms, Eqs.~\textup{\eqref{eq:Dnm-op}} and \textup{\eqref{eq:Dnm-ontic}}. The right panel records the general empirical preparation-property and its pointwise-solved ontic counterpart, Eqs.~\textup{\eqref{eq:generic-task}} and \textup{\eqref{eq:ontic-task-solved}}. In this way, the figure makes visually explicit the strengthening summarized abstractly in Eq.~\textup{\eqref{eq:hierarchy-core}}: one moves from preserving indistinguishability, to preserving distinguishability, to preserving the full empirical preparation profile of a finite preparation set.}
\label{fig:ec-hierarchy}
\end{figure}
Figure~\ref{fig:ec-hierarchy} gives the corresponding preparation-side picture: from exact preservation of indistinguishability, to exact preservation of distinguishability, to exact preservation of the full empirical preparation profile.

\subsection{Epistemic completeness}

We can now define the main notion of the paper. Because the manuscript works simultaneously with ontic extensions, ontological models, and operational theories, it is useful to separate three levels of the definition from the outset. First, one may ask whether a given ontic extension already preserves empirical preparation-properties as not-fine-tuned ontic task-values. Second, one may ask the same question after imposing the additional structural assumptions that turn the ontic extension into an ontological model. Third, one may ask whether a given operational theory admits at least one such ontological completion. The next three definitions formalize these three levels in exactly that order.

\begin{definition}[Epistemically complete ontic extension]
An ontic extension $Q$ of an operational theory $\cO$ is epistemically complete if for every finite preparation set $\bP=(P_1,\dots,P_n)$ and every empirical preparation-property $S_n^{(\cO)}(\bP)$ of the form \eqref{eq:generic-task}, the corresponding not-fine-tuned ontic task-value satisfies
\begin{equation}
S_n^{(Q)}(\bP)=S_n^{(\cO)}(\bP),
\end{equation}
where
\begin{equation}
S_n^{(Q)}(\bP)
=
\max_{M\in\cM_{\cO}}
\max_{\xi\in\Xi_\Lambda}
\sum_{x,k',k}
c_{k'}^x
\int_\Lambda d\lambda\, \xi(k'|k,\lambda)\,q(k,\lambda|P_x,M).
\end{equation}
\end{definition}

\begin{definition}[Epistemically complete ontological model]
An ontological model is epistemically complete if it is epistemically complete as an ontic extension and also satisfies $\lambda$-mediation and measurement independence.
\end{definition}

\begin{definition}[Epistemically complete operational theory]
An operational theory is epistemically complete if it admits at least one epistemically complete ontological completion.
\end{definition}

The distinction between these three levels is important. The first is defined directly for ontic extensions. The second is the special case relevant to ontological models. The third is a property of the operational theory itself.

\subsection{Soundness of the ontic-extension definition}

We defined epistemic completeness first at the level of ontic extensions because that is the weakest hidden-variable framework needed for the present question. The next proposition shows that this is a genuine extension of the usual ontological-model notion, not a different notion. Once $\lambda$-mediation and measurement independence are imposed, the ontic-extension task-value reduces exactly to the corresponding unrestricted ontic quantity of the associated epistemic states. Hence epistemic completeness for ontic extensions recovers epistemic completeness for ontological models as a special case.

\begin{proposition}
\label{prop:soundness}
Assume an ontic extension satisfies $\lambda$-mediation and measurement independence. Then for every preparation set $\bP=(P_1,\dots,P_n)$ with associated epistemic states $\bmu=(\mu_{P_1},\dots,\mu_{P_n})$ and every coefficient family $\{c_k^x\}$,
\begin{equation}
S_n^{(Q)}(\bP)=S_n^{(\Lambda)}(\bmu).
\end{equation}
\end{proposition}

\begin{proof}
We begin from the ontic-extension task-value:
\begin{equation}
S_n^{(Q)}(\bP)
=
\max_{M}
\max_{\xi(k'|k,\lambda)}
\sum_{x,k',k}
c_{k'}^x
\int_\Lambda d\lambda\,
\xi(k'|k,\lambda)\, q(k,\lambda|P_x,M).
\end{equation}
Here the second maximization is over all post-processings of the operational outcome \(k\) together with the ontic state \(\lambda\).

Now impose $\lambda$-mediation and measurement independence. These give
\begin{equation}
q(k,\lambda|P_x,M)=\mu_{P_x}(\lambda)\,\xi_M(k|\lambda),
\end{equation}
where \(\mu_{P_x}\) is the epistemic state associated with \(P_x\), and \(\xi_M(k|\lambda)\) is the response function of the measurement \(M\). Substituting this factorization into the previous expression yields
\begin{equation}
S_n^{(Q)}(\bP)
=
\max_M
\max_{\xi(k'|k,\lambda)}
\sum_{x,k',k}
c_{k'}^x
\int_\Lambda d\lambda\,
\xi(k'|k,\lambda)\,\mu_{P_x}(\lambda)\,\xi_M(k|\lambda).
\end{equation}

At this point, for fixed \(M\) and fixed \(\lambda\), the inner optimization over \(\xi(k'|k,\lambda)\) is again pointwise and linear. Since \(\sum_k \xi_M(k|\lambda)=1\), the dependence on the operational outcome \(k\) disappears after summation, and the optimal value of the post-processing is simply obtained by choosing the output \(k'\) that maximizes
\begin{equation}
\sum_x c_{k'}^x\,\mu_{P_x}(\lambda).
\end{equation}
Hence the value of the inner optimization is
\begin{equation}
\max_{k'}\sum_x c_{k'}^x\,\mu_{P_x}(\lambda),
\end{equation}
independently of \(M\). Therefore
\begin{equation}
S_n^{(Q)}(\bP)
=
\int_\Lambda d\lambda\,
\max_{k'}\sum_x c_{k'}^x\,\mu_{P_x}(\lambda).
\end{equation}
But this is exactly the expression for the unrestricted ontic quantity derived in Lemma~\ref{lem:ontic-solved}. Thus
\begin{equation}
S_n^{(Q)}(\bP)=S_n^{(\Lambda)}(\bmu),
\end{equation}
as claimed.
\end{proof}

Proposition~\ref{prop:soundness} shows that the ontic-extension definition of epistemic completeness is sound in the following precise sense: when the usual causal assumptions of ontological models are imposed, the extension-level notion reduces exactly to the ontological-model notion. In particular, any ontic extension that is epistemically complete and also satisfies $\lambda$-mediation and measurement independence yields an epistemically complete ontological model.
\subsection{Two epistemically complete comparison classes}

Before turning to the equality witness, it is useful to record two basic classes of operational theories in which epistemic completeness holds exactly. The first is unrestricted classical communication, which provides the natural classical benchmark for the present framework. The second is the class of finite-dimensional commuting quantum theories, which identifies a maximal classical sector inside quantum theory itself.

\begin{proposition}
\label{prop:classical-ec}
Unrestricted classical communication models are epistemically complete.
\end{proposition}

\begin{proof}
In a classical communication model, the transmitted classical message \(\omega\) can itself be taken as the ontic state. The preparation associated with input \(x\) is the probability distribution \(e_x(\omega)\), and the operational decoder already acts directly on the classical message through stochastic response functions \(d_M(k|\omega)\). Since every normalized conditional distribution over outputs given \(\omega\) is itself a valid classical decoder, the operationally realizable response schemes already coincide with the full set of all normalized response schemes on the ontic state space. Therefore, for every task and every preparation set, the operational task-value agrees exactly with the corresponding not-fine-tuned ontic quantity. Hence unrestricted classical communication models are epistemically complete.
\end{proof}

This first example is conceptually central. It shows that deviations from the equality witness cannot occur in unrestricted classical communication theories. Consequently, whenever quantum theory violates the witness, that violation already points toward a communication advantage over classical resources under an appropriate benchmark constraint.

A second positive comparison class arises from within quantum theory itself.
Here and below, by a finite-dimensional commuting quantum theory we mean the quantum theory associated with a fixed finite-dimensional commuting \(*\)-subalgebra of \(\mathcal B(\mathcal H)\), with preparations given by all density operators in that algebra and measurements given by all POVMs whose effects lie in that algebra.
\begin{proposition}
\label{prop:commuting-ec}
Every finite-dimensional commuting quantum theory is epistemically complete.
\end{proposition}

\begin{proof}
In a finite-dimensional commuting quantum theory, all allowed preparation states and all allowed measurement effects belong to a common commuting \(*\)-subalgebra of \(\mathcal B(\mathcal H)\). Hence there exists an orthonormal basis \(\{\ket{i}\}_{i=1}^d\) in which every allowed preparation state \(\rho_x\) and every allowed measurement effect \(M_k\) is diagonal:
\begin{equation}
\rho_x=\sum_{i=1}^d p_x(i)\,\dyad{i},
\qquad
M_k=\sum_{i=1}^d m_k(i)\,\dyad{i},
\end{equation}
with \(p_x(i)\ge 0\), \(\sum_i p_x(i)=1\), \(m_k(i)\ge 0\), and \(\sum_k m_k(i)=1\) for every \(i\). The operational statistics therefore take the classical stochastic form
\begin{equation}
p(k|\rho_x,M)=\Tr(\rho_x M_k)=\sum_{i=1}^d p_x(i)\,m_k(i).
\end{equation}
Thus the theory is operationally equivalent to a classical prepare-and-measure theory with underlying classical variable \(i\). Because the measurement optimization in the commuting quantum theory ranges over all POVMs in the common commuting algebra, it ranges exactly over all stochastic decoders \(m_k(i)\) on that classical variable. Therefore, for every task and every preparation set, the operational task-value agrees exactly with the corresponding unrestricted ontic quantity. Hence the theory is epistemically complete.
\end{proof}

This second example is equally important conceptually. It identifies a maximal classical sector inside finite-dimensional quantum theory itself: as long as one remains within a finite-dimensional commuting quantum theory, the preparation-side operational description is already complete in the present sense. The equality witness therefore becomes nontrivial only when genuinely noncommuting quantum structure enters.

Having identified these two epistemically complete comparison classes, we next isolate the particular family of tasks from which the equality witness is built. 

\section{The family of set-distinguishability tasks and the two averaged quantities}

Before proving the equality witness, it is useful to isolate the canonical family of empirical preparation-properties from which that witness is built. The point of the present section is not to introduce an additional benchmark beside distinguishability, but to make clear why one specific family of communication tasks already captures the structure needed for the central theorem. The equality witness will arise by comparing two natural averages extracted from this family, and its preparation-side significance becomes clearer once those quantities are viewed in that common operational setting.

\subsection{A canonical family of tasks}

The family \(\{D_{n,m}\}_{m=1}^{n-1}\) consists of the set-distinguishability tasks for a fixed finite set of \(n\) preparations. Each value \(D_{n,m}\) asks how well the same preparation set supports a different communication task: Bob must output an \(m\)-element set that contains the true preparation label.

The endpoint cases recover familiar operational quantities. The value \(D_{n,1}\) is ordinary distinguishability, since Bob must identify the label exactly. The value \(D_{n,n-1}\) is antidistinguishability, since Bob must exclude one label correctly. Intermediate values interpolate between these two extremes by asking how well the communicated preparation allows one to narrow the label down to a set of prescribed size. In this way, the family \(\{D_{n,m}\}_{m=1}^{n-1}\) gives a canonical collection of empirical preparation-properties naturally associated with the same finite set of preparations.

This viewpoint also clarifies the relation to informationally restricted prepare-and-measure frameworks. There, one studies correlations under bounds on the information content of the communicated ensemble, most notably through the one-shot accessible information \cite{Tavakoli2020informationally,Tavakoli2022informationally}. More recent work has sharpened this picture by showing that a broad class of natural physical assumptions can be organized through a common relaxation to one-shot accessible information, and by correspondingly identifying the optimal state-discrimination probability of the ensemble as a particularly central operational quantity \cite{PauwelsPironioTavakoli2025}. The present framework shares that operational language, but uses it for a different purpose. Since distinguishability and related guessing probabilities are themselves communication-task performance measures of the communicated preparations, it is natural to ask what exact constraints complete classical explanation imposes when one considers the full family of empirical communication-task quantities associated with a finite preparation set.

\subsection{From single-task preservation to family-wise preservation}

Since distinguishability is itself one empirical preparation-property, any notion that preserves all empirical communication-task quantities must in particular preserve distinguishability. The converse, however, is false: preserving distinguishability alone does not guarantee preservation of the remaining set-distinguishability quantities.

This comparison can be stated precisely. Bounded ontological distinctness on preparations preserves one distinguished member of the family, namely exact label discrimination \cite{Chaturvedi2020quantum}. Epistemic completeness, by contrast, demands exact preservation of the whole class \eqref{eq:generic-task}. The equality witness derived below is therefore not a reformulation of a single-task principle. Rather, it is the signature of a stronger preparation-side requirement: exact preservation of a whole family of empirical communication-task quantities associated with the same finite set of preparations.

With this canonical family now isolated, we are ready to prove the central theorem. The next section shows that epistemic completeness forces an exact equality between two natural averages extracted from these set-distinguishability quantities.

\section{Equality witness for epistemically complete theories}

We now come to the central theorem of the manuscript. The aim of this section is to show that epistemic completeness forces an exact equality between two different averages extracted from the same preparation profile. The important point is that this equality is not a contingent feature of quantum theory, classical communication, or any special ontological ansatz. It is forced by the structure of unrestricted ontic decoding itself.

For that reason, the proof proceeds in two stages. First, we identify an elementary identity for ordered lists of real numbers. Second, we translate that identity pointwise to epistemic states and then port it to the operational level using epistemic completeness. Presenting the argument in this order makes clear why the witness is an equality rather than an inequality: once ontic decoding is unrestricted, both averages are simply two reorganizations of the same pointwise optimization data.

\subsection{The underlying real-number identity}

The proof of the equality witness rests on a simple but powerful identity for ordered lists of real numbers.

\begin{lemma}
\label{lem:real-identity}
For every integer $n\ge2$ and every collection of real numbers $u_1,\dots,u_n$,
\begin{equation}
\label{eq:real-identity}
\sum_{1\le i<j\le n}\max\set{u_i,u_j}
=
\sum_{m=1}^{n-1}\ \max_{\substack{I\subset[n]\\ |I|=m}}\ \sum_{i\in I}u_i.
\end{equation}
\end{lemma}

\begin{proof}
Let $a_1\ge a_2\ge \cdots \ge a_n$ be the nonincreasing rearrangement of the numbers $u_1,\dots,u_n$. Then the left-hand side of \eqref{eq:real-identity} is
\begin{equation}
\sum_{x=1}^{n-1}(n-x)a_x,
\end{equation}
because $a_x$ is the larger entry in exactly $n-x$ unordered pairs.

On the right-hand side, for each fixed $m$, the maximum $m$-subset sum is obtained by choosing the $m$ largest entries, so
\begin{equation}
\max_{\substack{I\subset[n]\\ |I|=m}}\sum_{i\in I}u_i
=
\sum_{x=1}^m a_x.
\end{equation}
Therefore
\begin{equation}
\sum_{m=1}^{n-1}\max_{\substack{I\subset[n]\\ |I|=m}}\sum_{i\in I}u_i
=
\sum_{m=1}^{n-1}\sum_{x=1}^m a_x
=
\sum_{x=1}^{n-1}(n-x)a_x,
\end{equation}
which coincides with the left-hand side.
\end{proof}

The point of this identity is that it is purely combinatorial. It depends on no special property of quantum theory, no ontological-model assumption, and no optimization trick. Once translated pointwise to epistemic states, it yields the universal ontic equality.

\subsection{Ontic equality}

For any epistemic ensemble $\bmu=(\mu_1,\dots,\mu_n)$, the subset-identification task-values at the ontic level are
\begin{equation}
\label{eq:Dnm-ontic}
D_{n,m}^{(\Lambda)}(\bmu)
=
\frac1n
\int_\Lambda d\lambda\
\max_{\substack{I\subset[n]\\ |I|=m}}\ \sum_{i\in I}\mu_i(\lambda),
\end{equation}
which is the specialization of Eq.~\eqref{eq:ontic-task-solved} to the subset-identification task.

Similarly, for any pair $(i,j)$,
\begin{equation}
D_{2,1}^{(\Lambda)}(\mu_i,\mu_j)
=
\frac12
\int_\Lambda d\lambda\,
\max\set{\mu_i(\lambda),\mu_j(\lambda)}.
\end{equation}

Averaging these quantities yields
\begin{equation}
\avgD_n^{(\Lambda)}(\bmu)
=
\frac{1}{n(n-1)}
\int_\Lambda d\lambda
\sum_{i<j}\max\set{\mu_i(\lambda),\mu_j(\lambda)},
\end{equation}
and
\begin{equation}
\avgS_n^{(\Lambda)}(\bmu)
=
\frac{1}{n(n-1)}
\int_\Lambda d\lambda
\sum_{m=1}^{n-1}
\max_{\substack{I\subset[n]\\ |I|=m}}\sum_{i\in I}\mu_i(\lambda).
\end{equation}

Lemma~\ref{lem:real-identity} now implies the ontic equality immediately.

\begin{theorem}[Ontic equality]
\label{thm:ontic-equality}
For every finite epistemic ensemble $\bmu=(\mu_1,\dots,\mu_n)$,
\begin{equation}
\label{eq:ontic-equality}
\avgD_n^{(\Lambda)}(\bmu)=\avgS_n^{(\Lambda)}(\bmu).
\end{equation}
\end{theorem}

\begin{proof}
For every fixed ontic state $\lambda$, apply Lemma~\ref{lem:real-identity} to the real numbers $u_i=\mu_i(\lambda)$. This gives the pointwise identity
\begin{equation}
\sum_{i<j}\max\set{\mu_i(\lambda),\mu_j(\lambda)}
=
\sum_{m=1}^{n-1}
\max_{\substack{I\subset[n]\\ |I|=m}}
\sum_{i\in I}\mu_i(\lambda).
\end{equation}
Integrating over $\lambda$ and multiplying by $1/[n(n-1)]$ yields \eqref{eq:ontic-equality}.
\end{proof}

\subsection{Operational equality witness}

We can now port the ontic equality to the operational level.

\begin{theorem}[Equality witness for epistemic completeness]
\label{thm:equality-witness}
Let $\cO$ be an epistemically complete operational theory. Then for every finite preparation set $\bP=(P_1,\dots,P_n)$,
\begin{equation}
\label{eq:equality-witness}
\avgD_n^{(\cO)}(\bP)=\avgS_n^{(\cO)}(\bP).
\end{equation}
\end{theorem}

\begin{proof}
Fix an epistemically complete ontological completion of $\cO$ for the preparation set $\bP$, and let $\bmu=(\mu_1,\dots,\mu_n)$ be its associated epistemic ensemble. By epistemic completeness, the not-fine-tuned ontic task-values of this same ensemble reproduce every empirical preparation-property in the subset-identification family exactly. In particular,
\begin{equation}
D_{2,1}^{(\cO)}(P_i,P_j)=D_{2,1}^{(\Lambda)}(\mu_i,\mu_j)
\end{equation}
for every pair, and
\begin{equation}
D_{n,m}^{(\cO)}(\bP)=D_{n,m}^{(\Lambda)}(\bmu)
\end{equation}
for every $m\in\{1,\dots,n-1\}$. Hence
\begin{equation}
\avgD_n^{(\cO)}(\bP)=\avgD_n^{(\Lambda)}(\bmu)
\qquad\text{and}\qquad
\avgS_n^{(\cO)}(\bP)=\avgS_n^{(\Lambda)}(\bmu).
\end{equation}
The conclusion then follows from Theorem~\ref{thm:ontic-equality}.
\end{proof}

Theorem~\ref{thm:equality-witness} is the central result of the paper. It says that epistemic completeness imposes an exact accounting identity on preparation-side communication power. If one average extracted from the subset-identification profile is known, then the other is fixed.

\subsection{Deviation from the equality witness}

The natural measure of epistemic incompleteness is therefore
\begin{equation}
\label{eq:Delta-def}
\Delta_n^{(\cO)}(\bP)
:=
\avgD_n^{(\cO)}(\bP)-\avgS_n^{(\cO)}(\bP).
\end{equation}

A positive deviation means that the average pairwise discrimination power of the preparation set exceeds the average subset-identification power. A negative deviation means the opposite. Since epistemically complete theories must satisfy $\Delta_n^{(\cO)}(\bP)=0$ for every finite preparation set \bP, any nonzero value of $\Delta_n^{(\cO)}(\bP)$ witnesses epistemic incompleteness.

More is true: $\abs{\Delta_n^{(\cO)}(\bP)}$ lower-bounds the amount of hidden ontic task-power that every ontic completion must conceal. To see this, consider first the case $\Delta_n^{(\cO)}(\bP)>0$. Any operational decoder is, once conditioned on access to $\lambda$, a particular admissible ontic decoder. Hence for any ontic completion reproducing the operational pairwise distinguishabilities one has
\begin{equation}
\avgD_n^{(\Lambda)}(\bmu)\ge \avgD_n^{(\cO)}(\bP),
\end{equation}
and by Theorem~\ref{thm:ontic-equality},
\begin{equation}
\avgS_n^{(\Lambda)}(\bmu)=\avgD_n^{(\Lambda)}(\bmu)\ge \avgD_n^{(\cO)}(\bP)
=
\avgS_n^{(\cO)}(\bP)+\Delta_n^{(\cO)}(\bP).
\end{equation}
Hence the ontic average set distinguishability must exceed the operational one by at least $\Delta_n^{(\cO)}(\bP)$.

Similarly, if $\Delta_n^{(\cO)}(\bP)<0$, then any ontic completion reproducing the operational subset-identification profile must satisfy
\begin{equation}
\avgD_n^{(\Lambda)}(\bmu)=\avgS_n^{(\Lambda)}(\bmu)\ge \avgS_n^{(\cO)}(\bP)
=
\avgD_n^{(\cO)}(\bP)+\abs{\Delta_n^{(\cO)}(\bP)},
\end{equation}
so the ontic average pairwise distinguishability exceeds the operational one by at least $\abs{\Delta_n^{(\cO)}(\bP)}$.

Thus the equality witness is not only a yes-or-no test. Its magnitude measures the minimum hidden excess task-power that any ontic completion must hide.

The natural next question is whether quantum theory obeys this equality or departs from it. We therefore turn to explicit quantum ensembles and compute the resulting deviations from the witness.

\section{General implications of a nonzero deviation}

The equality witness has immediate quantitative, operational, and structural consequences. For a finite preparation set \(\bP=(P_1,\dots,P_n)\), define
\begin{equation}
\Delta_n^{(\cO)}(\bP):=\avgD_n^{(\cO)}(\bP)-\avgS_n^{(\cO)}(\bP).
\end{equation}
Any nonzero deviation \(\Delta_n^{(\cO)}(\bP)\neq 0\) already certifies epistemic incompleteness. More precisely, its magnitude lower-bounds the amount of hidden ontic communication power that every ontic or hidden-variable completion must conceal, while its sign identifies where that hidden excess must appear. In this sense, the equality witness is not merely qualitative: it is an accounting principle for hidden ontic task-power. The same nonzero deviation also has operational and structural implications, which we now make explicit.

\subsection{Benchmark-constrained communication advantage}

Because unrestricted classical communication models are epistemically complete, they satisfy the equality witness exactly for every finite classical preparation set. Therefore any preparation set with nonzero deviation must outperform unrestricted classical communication in an appropriately benchmark-constrained one-way communication task.

\subsubsection{Positive deviation}

Suppose
\begin{equation}
\Delta_n^{(\cO)}(\bP)=\avgD_n^{(\cO)}(\bP)-\avgS_n^{(\cO)}(\bP)>0.
\end{equation}
Consider the following communication task. Alice receives a label \(x\in[n]\), prepares \(P_x\), and sends it to Bob. Bob receives as side information a pair \(\{i,j\}\) promised to contain \(x\), and must guess whether \(x=i\) or \(x=j\). The optimal success probability of a strategy built from \(\bP\) is exactly the average pairwise distinguishability \(\avgD_n^{(\cO)}(\bP)\).

Now let \(\bC\) range over all preparation sets realizable by unrestricted classical communication, and impose the benchmark constraint
\begin{equation}
\avgS_n^{(\mathrm{cl})}(\bC)\le p.
\end{equation}
If we choose
\begin{equation}
p=\avgS_n^{(\cO)}(\bP),
\end{equation}
then the strategy built from \(\bP\) achieves
\begin{equation}
\avgD_n^{(\cO)}(\bP)=p+\Delta_n^{(\cO)}(\bP)>p.
\end{equation}
By contrast, every unrestricted classical communication preparation set \(\bC\) obeying \(\avgS_n^{(\mathrm{cl})}(\bC)\le p\) must satisfy
\begin{equation}
\avgD_n^{(\mathrm{cl})}(\bC)\le p,
\end{equation}
because unrestricted classical communication is epistemically complete and therefore satisfies
\begin{equation}
\avgD_n^{(\mathrm{cl})}(\bC)=\avgS_n^{(\mathrm{cl})}(\bC)
\end{equation}
for every finite classical preparation set \(\bC\). Hence the preparation set \(\bP\) yields a strict communication advantage over all unrestricted classical strategies under the same task-based benchmark.

\begin{figure}[t]
\centering
\includegraphics[width=0.82\linewidth]{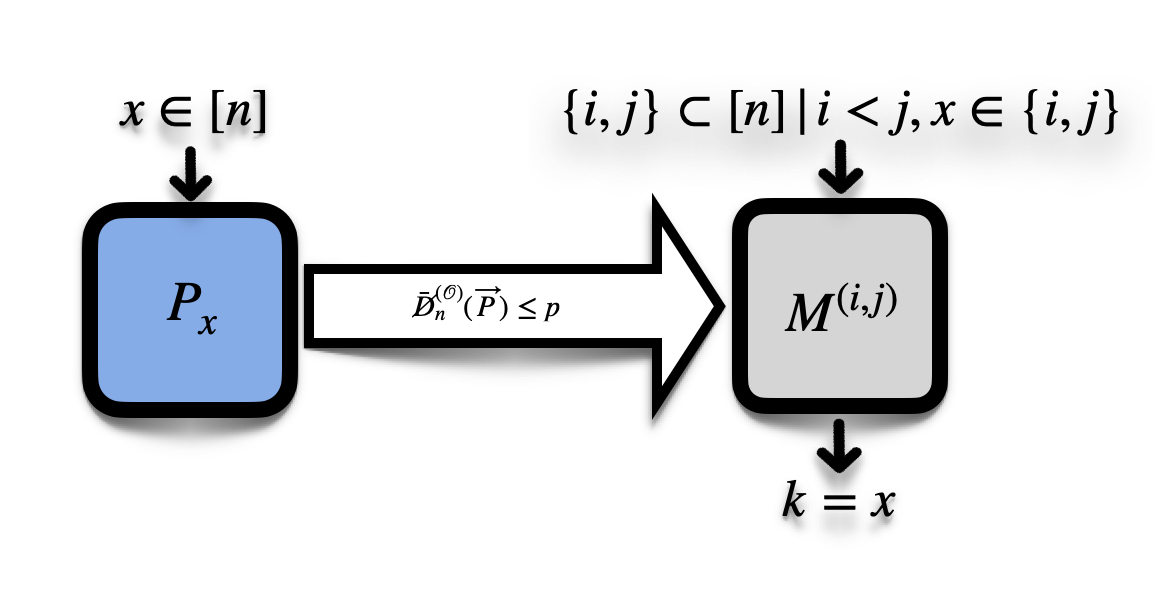}
\caption{Operational reading of the positive-deviation side of the equality witness. Alice receives \(x\in[n]\) and prepares \(P_x\). Bob receives a promised pair \(\{i,j\}\) containing \(x\) and must guess whether \(x=i\) or \(x=j\), so the task value is \(\avgD_n\). The benchmark is imposed on competing classical preparation sets \(\bC\) through \(\avgS_n^{(\mathrm{cl})}(\bC)\le p\), with \(p\) fixed by the given preparation set as \(p=\avgS_n^{(\cO)}(\bP)\). Because unrestricted classical communication obeys the equality witness exactly, any preparation set with \(\Delta_n^{(\cO)}(\bP)>0\) yields a communication advantage in this scenario.}
\label{fig:positive-task}
\end{figure}

\subsubsection{Negative deviation}

Suppose instead that
\begin{equation}
\Delta_n^{(\cO)}(\bP)<0,
\qquad\text{equivalently}\qquad
\avgS_n^{(\cO)}(\bP)>\avgD_n^{(\cO)}(\bP).
\end{equation}
Now let Bob's task be the averaged set-distinguishability task itself: after receiving the preparation, he is asked to output an \(m\)-element set containing the true label, with \(m\) chosen uniformly from \(\{1,\dots,n-1\}\). The optimal success probability of a strategy built from \(\bP\) is exactly \(\avgS_n^{(\cO)}(\bP)\).

Let \(\bC\) again range over all preparation sets realizable by unrestricted classical communication, and impose the benchmark constraint
\begin{equation}
\avgD_n^{(\mathrm{cl})}(\bC)\le p.
\end{equation}
Choosing
\begin{equation}
p=\avgD_n^{(\cO)}(\bP),
\end{equation}
the same preparation set achieves
\begin{equation}
\avgS_n^{(\cO)}(\bP)=p+\abs{\Delta_n^{(\cO)}(\bP)}>p.
\end{equation}
Again, every unrestricted classical communication preparation set \(\bC\) obeying \(\avgD_n^{(\mathrm{cl})}(\bC)\le p\) must satisfy
\begin{equation}
\avgS_n^{(\mathrm{cl})}(\bC)\le p,
\end{equation}
because unrestricted classical communication satisfies
\begin{equation}
\avgD_n^{(\mathrm{cl})}(\bC)=\avgS_n^{(\mathrm{cl})}(\bC)
\end{equation}
for every finite classical preparation set \(\bC\). Therefore a negative deviation also implies communication advantage.

\begin{figure}[t]
\centering
\includegraphics[width=0.82\linewidth]{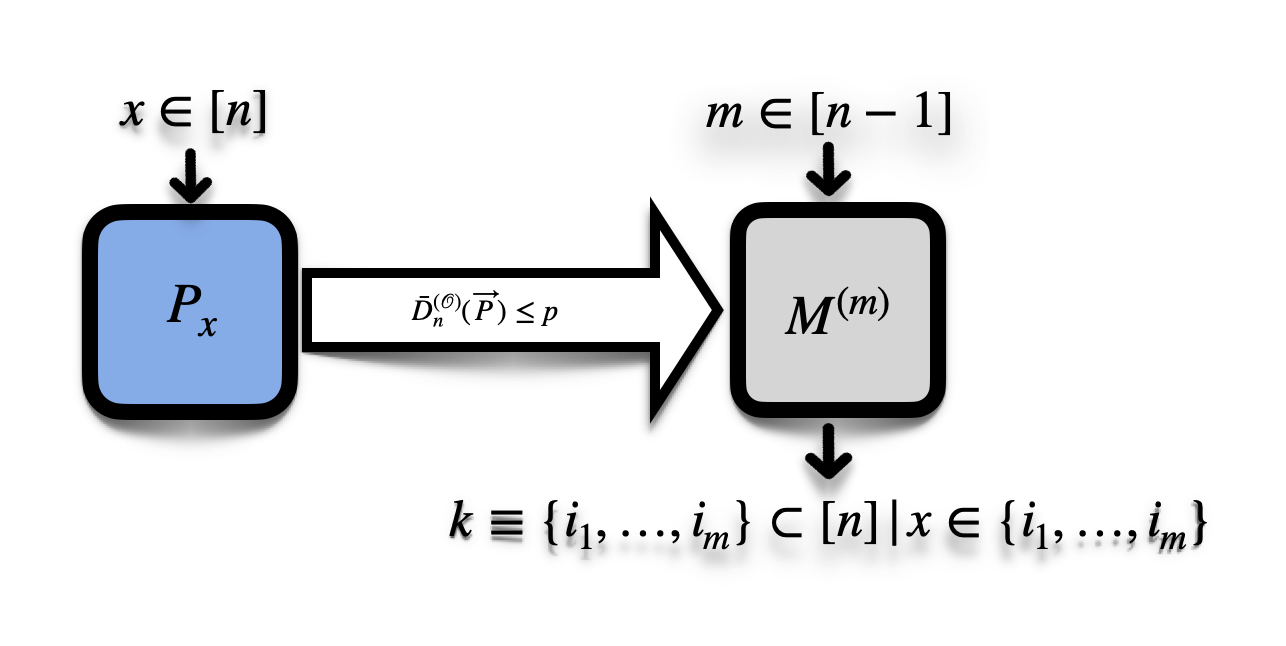}
\caption{Operational reading of the negative-deviation side of the equality witness. Alice receives \(x\in[n]\) and prepares \(P_x\). Bob receives \(m\in\{1,\dots,n-1\}\) and must output an \(m\)-element set containing the true label, so the task value is \(\avgS_n\). The benchmark is imposed on competing classical preparation sets \(\bC\) through \(\avgD_n^{(\mathrm{cl})}(\bC)\le p\), with \(p\) fixed by the given preparation set as \(p=\avgD_n^{(\cO)}(\bP)\). Because unrestricted classical communication obeys the equality witness exactly, any preparation set with \(\Delta_n^{(\cO)}(\bP)<0\) yields a communication advantage in this dual scenario.}
\label{fig:negative-task}
\end{figure}

\begin{theorem}[Communication advantage from any nonzero deviation]
\label{thm:comm-adv}
Every preparation set \(\bP\) with \(\Delta_n^{(\cO)}(\bP)\neq 0\) yields an advantage over unrestricted classical communication in a one-way communication task under a natural task-based benchmark constraint.
\end{theorem}

\begin{proof}
The positive- and negative-deviation constructions above cover the two possible signs of \(\Delta_n^{(\cO)}(\bP)\).
\end{proof}

\subsection{Coherence witness for preparation sets}

The same nonzero deviation has a preparation-side structural consequence.

\begin{proposition}[Nonzero deviation witnesses coherence]
\label{prop:coherence-witness}
Let \(\bP\) be a finite set of quantum preparations. If \(\bP\) can be jointly realized within a finite-dimensional commuting quantum theory, then
\begin{equation}
\Delta_n^{(\cQ)}(\bP)=0.
\end{equation}
Hence any nonzero deviation certifies that the preparation set cannot be jointly realized within any finite-dimensional commuting quantum theory.
\end{proposition}

\begin{proof}
This is immediate from Proposition~\ref{prop:commuting-ec} and Theorem~\ref{thm:equality-witness}. Every finite-dimensional commuting quantum theory is epistemically complete and therefore satisfies the equality witness exactly.
\end{proof}

In this precise preparation-side sense, any nonzero deviation certifies coherence of the preparation set.

\subsection{Measurement-incompatibility witness in any realizing implementation}

Finally, the same witness has a measurement-side implication for any implementation that realizes the certifying task values.

\begin{proposition}[Nonzero deviation witnesses measurement incompatibility]
\label{prop:incompatibility-witness}
Consider any concrete realization of the task values entering \(\avgD_n\) and \(\avgS_n\) by a finite family of measurements for a given finite preparation set. If that certifying family is jointly measurable, then the prepare-and-measure theory generated by the corresponding parent POVM and its classical postprocessings is operationally equivalent to a classical prepare-and-measure model and therefore satisfies
\begin{equation}
\avgD_n=\avgS_n.
\end{equation}
Consequently, any realization attaining a nonzero deviation must employ an incompatible certifying family of measurements.
\end{proposition}

\begin{proof}
If the certifying measurements are jointly measurable, there exists a parent POVM \(\{G_\lambda\}_\lambda\) together with classical postprocessings \(q(k|M,\lambda)\) such that every certifying measurement effect is obtained as a postprocessing of the parent. Hence every certifying statistic takes the form
\begin{equation}
p(k|P_x,M)=\sum_\lambda \Tr(\rho_x G_\lambda)\,q(k|M,\lambda).
\end{equation}
Since any classical postprocessing of an implementable parent POVM is again an implementable measurement, the theory generated by the parent POVM ranges over all stochastic decoders \(q(k|\lambda)\) on the classical parent outcome \(\lambda\). This is exactly a classical prepare-and-measure model with ontic state \(\lambda\). By Proposition~\ref{prop:classical-ec} and Theorem~\ref{thm:equality-witness}, every such model satisfies the equality exactly, i.e. the realized deviation vanishes. The claim follows by contraposition.
\end{proof}

Taken together, these consequences show that a single nonzero deviation simultaneously quantifies epistemic incompleteness, implies benchmark-constrained quantum communication advantage, certifies coherence of the preparations, and witnesses measurement incompatibility in any realization that attains it.

\section{Quantum violations and extremal geometry}

The equality theorem is exact and universal for epistemically complete theories. The next question is therefore immediate: does quantum theory satisfy it? This section answers that question in the negative. We first evaluate the canonical trine and tetrahedral qubit ensembles exactly, thereby obtaining explicit positive-direction violations. We then record opposite-sign quantum violations and describe the geometric picture in the \((\avgS_n,\avgD_n)\) plane. The extremality claims for \(n=3\) and \(n=4\) are then certified numerically in Sec.~\ref{CABIOQT}, where the SDP hierarchy and see-saw bounds are developed in full detail.

\subsection{Dual formulation for quantum set-distinguishability}

The exact evaluations below will use the semidefinite-programming dual of the quantum set-distinguishability task.

\begin{proposition}[Dual formulation]
\label{prop:subset-dual}
Let \(\brho=(\rho_1,\dots,\rho_n)\) be a finite quantum ensemble and fix \(m\in\{1,\dots,n-1\}\). For each subset \(I\subset[n]\) with \(|I|=m\), define
\begin{equation}
A_I:=\frac1n\sum_{i\in I}\rho_i.
\end{equation}
Then
\begin{equation}
D_{n,m}^{(\cQ)}(\brho)
=
\min_{K\succeq 0}\Bigl\{\Tr(K):K\succeq A_I\ \text{for all }I\subset[n],\ |I|=m\Bigr\}.
\end{equation}
\end{proposition}

\begin{proof}
The primal semidefinite program for \(D_{n,m}^{(\cQ)}(\brho)\) is
\begin{equation}
D_{n,m}^{(\cQ)}(\brho)
=
\max_{\{M_I\}}
\sum_{|I|=m}\Tr(A_I M_I),
\end{equation}
subject to \(M_I\succeq0\) and \(\sum_{|I|=m}M_I=\id\). The corresponding dual is exactly the stated minimization over a single positive semidefinite operator \(K\) dominating every \(A_I\). Strong duality holds by Slater's condition.
\end{proof}

\subsection{Trine ensemble}

We first consider the trine ensemble of pure qubit states. In Bloch-vector form, take
\begin{equation}
\mathbf n_1=(1,0,0),\quad
\mathbf n_2=\left(-\frac12,\frac{\sqrt3}{2},0\right),\quad
\mathbf n_3=\left(-\frac12,-\frac{\sqrt3}{2},0\right),
\end{equation}
and define
\begin{equation}
\rho_i=\frac12\left(\id+\mathbf n_i\cdot\boldsymbol\sigma\right),\qquad i=1,2,3.
\end{equation}

\begin{proposition}[Exact trine violation]
\label{prop:trine-exact}
For the trine ensemble \(\brho_\triangle=(\rho_1,\rho_2,\rho_3)\),
\begin{align}
\avgD_3^{(\cQ)}(\brho_\triangle)
&=\frac12\left(1+\frac{\sqrt3}{2}\right),\\
D_{3,1}^{(\cQ)}(\brho_\triangle)
&=\frac23,\\
D_{3,2}^{(\cQ)}(\brho_\triangle)
&=1,\\
\avgS_3^{(\cQ)}(\brho_\triangle)
&=\frac56,
\end{align}
and therefore
\begin{equation}
\Delta_3^{(\cQ)}(\brho_\triangle)
=
\frac12\left(1+\frac{\sqrt3}{2}\right)-\frac56
=
\frac{3\sqrt3-4}{12}
\approx 0.0997.
\end{equation}
\end{proposition}

\begin{proof}
Every pair of trine states has overlap squared \(\abs{\braket{\psi_i}{\psi_j}}^2=1/4\), so the Helstrom formula gives
\begin{equation}
D_{2,1}^{(\cQ)}(\rho_i,\rho_j)
=
\frac12\left(1+\sqrt{1-\frac14}\right)
=
\frac12\left(1+\frac{\sqrt3}{2}\right).
\end{equation}
Averaging over the three pairs yields
\begin{equation}
\avgD_3^{(\cQ)}(\brho_\triangle)=\frac12\left(1+\frac{\sqrt3}{2}\right).
\end{equation}

For \(D_{3,1}^{(\cQ)}\), the POVM
\begin{equation}
M_i=\frac23\rho_i,\qquad i=1,2,3,
\end{equation}
is feasible because \(\rho_1+\rho_2+\rho_3=\frac32\id\). Its success probability is
\begin{equation}
\frac13\sum_{i=1}^3\Tr(\rho_i M_i)=\frac23.
\end{equation}
On the dual side, the operators to dominate are \(A_i=\rho_i/3\). The choice
\begin{equation}
K=\frac13\id
\end{equation}
is feasible and has \(\Tr(K)=2/3\). Hence \(D_{3,1}^{(\cQ)}(\brho_\triangle)=2/3\).

For \(D_{3,2}^{(\cQ)}\), label the three outcomes by the missing singleton and define
\begin{equation}
N_i=\frac23\rho_i^\perp,\qquad i=1,2,3,
\end{equation}
where \(\rho_i^\perp=\id-\rho_i\). Since \(\rho_1^\perp+\rho_2^\perp+\rho_3^\perp=\frac32\id\), this is a POVM. Moreover, for each input \(i\), the wrongly excluded outcome \(i\) occurs with probability
\begin{equation}
\Tr(\rho_iN_i)=\frac23\Tr(\rho_i\rho_i^\perp)=0.
\end{equation}
Thus the task succeeds perfectly and \(D_{3,2}^{(\cQ)}(\brho_\triangle)=1\).

Therefore
\begin{equation}
\avgS_3^{(\cQ)}(\brho_\triangle)=\frac12\left(\frac23+1\right)=\frac56,
\end{equation}
and the stated expression for \(\Delta_3^{(\cQ)}(\brho_\triangle)\) follows.
\end{proof}

Thus the trine ensemble yields an explicit positive violation of the equality witness. Quantum theory is therefore epistemically incomplete already on this simplest nontrivial three-state qubit fragment.

\subsection{Tetrahedral ensemble}

We next consider the tetrahedral ensemble of pure qubit states, whose Bloch vectors point to the vertices of a regular tetrahedron:
\begin{equation}
\mathbf t_1=\frac{(1,1,1)}{\sqrt3},\quad
\mathbf t_2=\frac{(1,-1,-1)}{\sqrt3},\quad
\mathbf t_3=\frac{(-1,1,-1)}{\sqrt3},\quad
\mathbf t_4=\frac{(-1,-1,1)}{\sqrt3}.
\end{equation}
Set
\begin{equation}
\rho_i=\frac12\left(\id+\mathbf t_i\cdot\boldsymbol\sigma\right),\qquad i=1,2,3,4.
\end{equation}

\begin{proposition}[Exact tetrahedral violation]
\label{prop:tetra-exact}
For the tetrahedral ensemble \(\brho_{\mathrm{tet}}=(\rho_1,\rho_2,\rho_3,\rho_4)\),
\begin{align}
\avgD_4^{(\cQ)}(\brho_{\mathrm{tet}})
&=\frac12\left(1+\sqrt{\frac23}\right),\\
D_{4,1}^{(\cQ)}(\brho_{\mathrm{tet}})
&=\frac12,\\
D_{4,2}^{(\cQ)}(\brho_{\mathrm{tet}})
&=\frac12+\frac{1}{2\sqrt3},\\
D_{4,3}^{(\cQ)}(\brho_{\mathrm{tet}})
&=1,\\
\avgS_4^{(\cQ)}(\brho_{\mathrm{tet}})
&=\frac23+\frac{1}{6\sqrt3},
\end{align}
and therefore
\begin{equation}
\Delta_4^{(\cQ)}(\brho_{\mathrm{tet}})
=
\frac12\left(1+\sqrt{\frac23}\right)-\frac23-\frac{1}{6\sqrt3}
\approx 0.14535658.
\end{equation}
\end{proposition}

\begin{proof}
Any two tetrahedral states satisfy \(\mathbf t_i\cdot\mathbf t_j=-1/3\), hence
\begin{equation}
\abs{\braket{\psi_i}{\psi_j}}^2=\frac{1+\mathbf t_i\cdot\mathbf t_j}{2}=\frac13.
\end{equation}
The Helstrom formula then gives
\begin{equation}
D_{2,1}^{(\cQ)}(\rho_i,\rho_j)
=
\frac12\left(1+\sqrt{1-\frac13}\right)
=
\frac12\left(1+\sqrt{\frac23}\right),
\end{equation}
and therefore
\begin{equation}
\avgD_4^{(\cQ)}(\brho_{\mathrm{tet}})
=
\frac12\left(1+\sqrt{\frac23}\right).
\end{equation}

For \(D_{4,1}^{(\cQ)}\), the tetrahedral POVM
\begin{equation}
M_i=\frac12\rho_i,\qquad i=1,2,3,4,
\end{equation}
is feasible because \(\sum_i\rho_i=2\id\). Its success probability is
\begin{equation}
\frac14\sum_{i=1}^4\Tr(\rho_iM_i)=\frac12.
\end{equation}
On the dual side, \(A_i=\rho_i/4\), so \(K=\frac14\id\) is feasible with \(\Tr(K)=1/2\). Hence \(D_{4,1}^{(\cQ)}(\brho_{\mathrm{tet}})=1/2\).

For \(D_{4,2}^{(\cQ)}\), for each pair \(I=\{i,j\}\) let \(\mathbf n_I\) be the unit Bloch vector parallel to \(\mathbf t_i+\mathbf t_j\). These are the six octahedral directions. Define
\begin{equation}
M_I=\frac13P_{\mathbf n_I},
\qquad
P_{\mathbf n_I}=\frac12(\id+\mathbf n_I\cdot\boldsymbol\sigma).
\end{equation}
The six octahedral projectors sum to \(3\id\), so this is a POVM. If \(i\in I\), then
\begin{equation}
\mathbf t_i\cdot\mathbf n_I=\frac{1}{\sqrt3},
\end{equation}
hence
\begin{equation}
\Tr(\rho_iP_{\mathbf n_I})=\frac12\left(1+\frac1{\sqrt3}\right).
\end{equation}
Each input belongs to exactly three pairs, so
\begin{equation}
D_{4,2}^{(\cQ)}(\brho_{\mathrm{tet}})
=
\frac14\cdot 4\cdot 3\cdot \frac13\cdot \frac12\left(1+\frac1{\sqrt3}\right)
=
\frac12+\frac{1}{2\sqrt3}.
\end{equation}
On the dual side, the operator \(A_I=(\rho_i+\rho_j)/4\) has largest eigenvalue \((1+1/\sqrt3)/4\), so
\begin{equation}
K=\frac14\left(1+\frac1{\sqrt3}\right)\id
\end{equation}
is feasible and has the same trace.

For \(D_{4,3}^{(\cQ)}\), label the four outcomes by the missing singleton and define
\begin{equation}
N_i=\frac12\rho_i^\perp,\qquad i=1,2,3,4.
\end{equation}
Since \(\sum_i\rho_i^\perp=2\id\), this is a POVM. Moreover \(\Tr(\rho_iN_i)=0\) for every \(i\), so the true label is never excluded and \(D_{4,3}^{(\cQ)}(\brho_{\mathrm{tet}})=1\).

Therefore
\begin{equation}
\avgS_4^{(\cQ)}(\brho_{\mathrm{tet}})
=
\frac13\left(\frac12+\frac12+\frac{1}{2\sqrt3}+1\right)
=
\frac23+\frac{1}{6\sqrt3},
\end{equation}
and the stated formula for \(\Delta_4^{(\cQ)}(\brho_{\mathrm{tet}})\) follows.
\end{proof}

The tetrahedral ensemble therefore yields a larger positive violation than the trine ensemble. Its extremality for the \(n=4\) equality will be certified numerically in Sec.~\ref{CABIOQT}.

\subsection{Negative-direction violations}

The equality witness is genuinely bidirectional. In particular, there exist quantum preparation sets for which
\begin{equation}
\avgS_n^{(\cQ)}(\bP)>\avgD_n^{(\cQ)}(\bP).
\end{equation}
Such examples do not arise among the canonical pure-state qubit ensembles above, but they do arise for mixed-state ensembles in higher dimension. In the three-preparation case, numerical optimization over mixed qutrit ensembles yields an opposite-sign deviation of approximately
\begin{equation}
\avgS_3^{(\cQ)}-\avgD_3^{(\cQ)}\approx 0.0277.
\end{equation}

The importance of this observation is conceptual. It shows that quantum theory fails the equality witness in both directions: not only can pairwise discrimination exceed what average set-distinguishability would allow in an epistemically complete theory, but the reverse imbalance can also occur.

\subsection{Geometric picture in the \((\avgS_3,\avgD_3)\) plane}

For \(n=3\), the equality witness can be visualized directly in the \((\avgS_3,\avgD_3)\) plane. The exact trine calculation determines an explicit positive-direction quantum point, while the mixed-state qutrit example determines an explicit opposite-sign point.

\begin{figure}[t]
\centering
\includegraphics[width=0.72\linewidth]{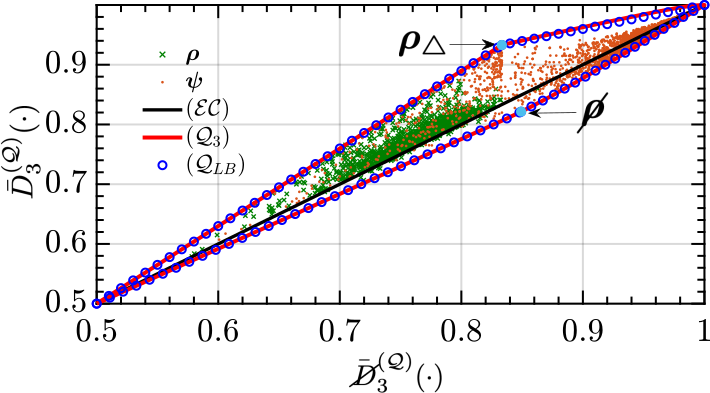}
\caption{Three-preparation geometry of the equality witness in the \((\avgS_3,\avgD_3)\) plane. The diagonal line \(\avgD_3=\avgS_3\) is the exact locus satisfied by every epistemically complete theory. The trine ensemble lies above the line and certifies a positive deviation. The opposite-sign mixed-state qutrit example lies below the line and certifies a negative deviation. The visibility and leakage families introduced in the next section trace explicit positive-direction nonclassical paths starting from the trine point.}
\label{fig:main-region}
\end{figure}

Figure~\ref{fig:main-region} makes the bidirectionality of the witness immediate: the canonical trine family lies above the equality line, while the mixed-state qutrit example lies below it. In particular, the upper boundary point at
\begin{equation}
(\avgS_3,\avgD_3)=\left(\frac56,\frac12\left(1+\frac{\sqrt3}{2}\right)\right)
\end{equation}
is attained by the trine ensemble and, as shown numerically in Sec.~\ref{CABIOQT}, is the exact positive-direction optimum for the three-preparation equality.

\subsection{Interpretation of the quantum violations}

The violated equality has a direct physical meaning. In an epistemically complete theory, the average pairwise distinguishability and the average set-distinguishability of a finite preparation set must agree because they arise as two exact operational shadows of the same unrestricted ontic quantities. In quantum theory they do not. Therefore any ontological completion of the quantum formalism must contain additional hidden communication power that is not operationally accessible. The sign of the deviation identifies where that hidden excess must appear, and its magnitude quantifies how much excess must be concealed.

At this stage one knows that quantum theory violates the witness in both directions, and that the canonical trine and tetrahedral ensembles already furnish exact positive-direction violations. The next step is to determine whether these violations are confined to the ideal symmetric points or persist throughout physically natural deformations of the same ensembles. We now show that they do.

\section{Visibility and leakage robustness}

The exact trine and tetrahedral points are canonical positive-direction violators of the equality witness. The purpose of the present section is to show that these violations are not isolated symmetric accidents. We study two explicit continuations of the canonical qubit ensembles: visibility families obtained by depolarizing the preparations, and leakage families obtained by adjoining an orthogonal branch that reveals the preparation label. In both cases, the positive-direction violation survives throughout the full nontrivial parameter range.

\subsection{Visibility families}

For a pure-state ensemble \(\brho=(\rho_1,\dots,\rho_n)\), define the corresponding visibility family by
\begin{equation}
\label{eq:visibility-family}
\rho_i^{(v)}
=
v\,\rho_i+(1-v)\frac{\id}{2},
\qquad
v\in[0,1].
\end{equation}

The theorem below evaluates the relevant quantities exactly for the trine and tetrahedral families. The proof follows the same pattern in each case: the pairwise terms are obtained from the Helstrom formula, while the set-distinguishability terms are certified by matching explicit POVMs with explicit dual feasible operators.

\begin{theorem}[Visibility robustness]
\label{thm:visibility}
For the visibility-deformed trine family,
\begin{align}
\avgD_3^{(\cQ)}(\brho_\triangle^{(v)})
&=
\frac12+\frac{\sqrt3}{4}v,\\
D_{3,1}^{(\cQ)}(\brho_\triangle^{(v)})
&=
\frac{1+v}{3},\\
D_{3,2}^{(\cQ)}(\brho_\triangle^{(v)})
&=
\frac{2+v}{3},\\
\avgS_3^{(\cQ)}(\brho_\triangle^{(v)})
&=
\frac12+\frac{v}{3},
\end{align}
and therefore
\begin{equation}
\Delta_3^{(\cQ)}(\brho_\triangle^{(v)})
=
v\,\frac{3\sqrt3-4}{12}.
\end{equation}

For the visibility-deformed tetrahedral family,
\begin{align}
\avgD_4^{(\cQ)}(\brho_{\mathrm{tet}}^{(v)})
&=
\frac12+\frac{\sqrt6}{6}v,\\
D_{4,1}^{(\cQ)}(\brho_{\mathrm{tet}}^{(v)})
&=
\frac{1+v}{4},\\
D_{4,2}^{(\cQ)}(\brho_{\mathrm{tet}}^{(v)})
&=
\frac12+\frac{v}{2\sqrt3},\\
D_{4,3}^{(\cQ)}(\brho_{\mathrm{tet}}^{(v)})
&=
\frac{3+v}{4},\\
\avgS_4^{(\cQ)}(\brho_{\mathrm{tet}}^{(v)})
&=
\frac12+\frac{1+1/\sqrt3}{6}\,v,
\end{align}
and therefore
\begin{equation}
\Delta_4^{(\cQ)}(\brho_{\mathrm{tet}}^{(v)})
=
v\left(
\frac{\sqrt6}{6}
-\frac16
-\frac{1}{6\sqrt3}
\right).
\end{equation}
In particular, both canonical families violate the equality witness for every \(v>0\).
\end{theorem}

\begin{proof}
We begin with the pairwise terms. Depolarization rescales Bloch-vector differences by the factor \(v\), so the Helstrom expression immediately yields
\begin{equation}
\avgD_3^{(\cQ)}(\brho_\triangle^{(v)})=\frac12+\frac{\sqrt3}{4}v,
\qquad
\avgD_4^{(\cQ)}(\brho_{\mathrm{tet}}^{(v)})=\frac12+\frac{\sqrt6}{6}v.
\end{equation}

We now turn to the trine family. For \(m=1\), the same trine POVM as in the pure-state case,
\begin{equation}
M_i=\frac23\rho_i,\qquad i=1,2,3,
\end{equation}
remains feasible. Its success probability on the visibility-deformed family is
\begin{align}
\frac13\sum_{i=1}^3 \Tr(\rho_i^{(v)}M_i)
&=
\frac13\sum_{i=1}^3 \Tr\!\left(\Bigl(v\rho_i+(1-v)\frac{\id}{2}\Bigr)\frac23\rho_i\right)\nonumber\\
&=
\frac13\sum_{i=1}^3 \frac23\left(v+\frac{1-v}{2}\right)
=
\frac{1+v}{3}.
\end{align}
For dual certification, the relevant operators are
\begin{equation}
A_i=\frac13\rho_i^{(v)}.
\end{equation}
Since the largest eigenvalue of \(\rho_i^{(v)}\) is \((1+v)/2\), the operator
\begin{equation}
K=\frac{1+v}{6}\id
\end{equation}
satisfies \(K\succeq A_i\) for every \(i\), and
\begin{equation}
\Tr(K)=\frac{1+v}{3}.
\end{equation}
Hence
\begin{equation}
D_{3,1}^{(\cQ)}(\brho_\triangle^{(v)})=\frac{1+v}{3}.
\end{equation}

For \(m=2\), use the same exclusion POVM as in the pure-state trine case,
\begin{equation}
N_i=\frac23\rho_i^\perp,\qquad i=1,2,3.
\end{equation}
The task succeeds unless the omitted singleton equals the true input. For input \(i\), the failure probability is
\begin{align}
\Tr(\rho_i^{(v)}N_i)
&=
\Tr\!\left(\Bigl(v\rho_i+(1-v)\frac{\id}{2}\Bigr)\frac23\rho_i^\perp\right)\nonumber\\
&=
\frac23\left(0+\frac{1-v}{2}\right)
=
\frac{1-v}{3}.
\end{align}
Therefore the success probability is
\begin{equation}
D_{3,2}^{(\cQ)}(\brho_\triangle^{(v)})
=
1-\frac{1-v}{3}
=
\frac{2+v}{3}.
\end{equation}
On the dual side, for each two-element set \(I\),
\begin{equation}
A_I=\frac13\sum_{i\in I}\rho_i^{(v)}.
\end{equation}
For a trine pair, the corresponding Bloch-vector sum has norm \(1\), so the largest eigenvalue of \(\sum_{i\in I}\rho_i^{(v)}\) is \(1+v/2\). Hence the largest eigenvalue of \(A_I\) is \((2+v)/6\), and
\begin{equation}
K=\frac{2+v}{6}\id
\end{equation}
is dual feasible with
\begin{equation}
\Tr(K)=\frac{2+v}{3}.
\end{equation}
Thus
\begin{equation}
D_{3,2}^{(\cQ)}(\brho_\triangle^{(v)})=\frac{2+v}{3}.
\end{equation}
Averaging \(D_{3,1}^{(\cQ)}\) and \(D_{3,2}^{(\cQ)}\) gives
\begin{equation}
\avgS_3^{(\cQ)}(\brho_\triangle^{(v)})=\frac12+\frac{v}{3},
\end{equation}
and hence the stated expression for \(\Delta_3^{(\cQ)}(\brho_\triangle^{(v)})\).

We now treat the tetrahedral family. For \(m=1\), the tetrahedral POVM
\begin{equation}
M_i=\frac12\rho_i,\qquad i=1,2,3,4,
\end{equation}
remains feasible. Its success probability is
\begin{align}
\frac14\sum_{i=1}^4 \Tr(\rho_i^{(v)}M_i)
&=
\frac14\sum_{i=1}^4 \Tr\!\left(\Bigl(v\rho_i+(1-v)\frac{\id}{2}\Bigr)\frac12\rho_i\right)\nonumber\\
&=
\frac14\sum_{i=1}^4 \frac12\left(v+\frac{1-v}{2}\right)
=
\frac{1+v}{4}.
\end{align}
The dual operators are \(A_i=\rho_i^{(v)}/4\), whose largest eigenvalue is \((1+v)/8\), so
\begin{equation}
K=\frac{1+v}{8}\id
\end{equation}
is feasible and has
\begin{equation}
\Tr(K)=\frac{1+v}{4}.
\end{equation}
Hence
\begin{equation}
D_{4,1}^{(\cQ)}(\brho_{\mathrm{tet}}^{(v)})=\frac{1+v}{4}.
\end{equation}

For \(m=2\), use the same octahedral POVM as in the pure tetrahedral case. For each pair \(I=\{i,j\}\), let \(\mathbf n_I\) be the unit vector parallel to \(\mathbf t_i+\mathbf t_j\), and set
\begin{equation}
M_I=\frac13 P_{\mathbf n_I},
\qquad
P_{\mathbf n_I}=\frac12(\id+\mathbf n_I\cdot\boldsymbol\sigma).
\end{equation}
As before, the six projectors sum to \(3\id\), so this is a POVM. If \(i\in I\), then
\begin{equation}
\mathbf t_i\cdot \mathbf n_I=\frac1{\sqrt3},
\end{equation}
and therefore
\begin{equation}
\Tr(\rho_i^{(v)}P_{\mathbf n_I})
=
\frac12\left(1+\frac{v}{\sqrt3}\right).
\end{equation}
Each input belongs to exactly three pairs, so
\begin{align}
D_{4,2}^{(\cQ)}(\brho_{\mathrm{tet}}^{(v)})
&=
\frac14\sum_{i=1}^4\sum_{I\ni i}\Tr(\rho_i^{(v)}M_I)\nonumber\\
&=
\frac14\cdot 4\cdot 3\cdot \frac13\cdot \frac12\left(1+\frac{v}{\sqrt3}\right)
=
\frac12+\frac{v}{2\sqrt3}.
\end{align}
For the dual, the relevant operators are
\begin{equation}
A_I=\frac14\sum_{i\in I}\rho_i^{(v)}.
\end{equation}
For a tetrahedral pair, \(\|\mathbf t_i+\mathbf t_j\|=2/\sqrt3\), so the largest eigenvalue of \(A_I\) is \((1+v/\sqrt3)/4\). Hence
\begin{equation}
K=\frac14\left(1+\frac{v}{\sqrt3}\right)\id
\end{equation}
is feasible, with
\begin{equation}
\Tr(K)=\frac12+\frac{v}{2\sqrt3}.
\end{equation}
Therefore
\begin{equation}
D_{4,2}^{(\cQ)}(\brho_{\mathrm{tet}}^{(v)})=\frac12+\frac{v}{2\sqrt3}.
\end{equation}

For \(m=3\), use the same exclusion POVM
\begin{equation}
N_i=\frac12\rho_i^\perp,\qquad i=1,2,3,4.
\end{equation}
For input \(i\), the failure probability is
\begin{align}
\Tr(\rho_i^{(v)}N_i)
&=
\Tr\!\left(\Bigl(v\rho_i+(1-v)\frac{\id}{2}\Bigr)\frac12\rho_i^\perp\right)\nonumber\\
&=
\frac12\left(0+\frac{1-v}{2}\right)
=
\frac{1-v}{4},
\end{align}
so the success probability is
\begin{equation}
D_{4,3}^{(\cQ)}(\brho_{\mathrm{tet}}^{(v)})
=
1-\frac{1-v}{4}
=
\frac{3+v}{4}.
\end{equation}
On the dual side, for each three-element set \(I\),
\begin{equation}
A_I=\frac14\sum_{i\in I}\rho_i^{(v)}.
\end{equation}
Using \(\sum_{i=1}^4\rho_i^{(v)}=2\id\), we have
\begin{equation}
\sum_{i\in I}\rho_i^{(v)}=2\id-\rho_k^{(v)}
\end{equation}
for the omitted label \(k\). The largest eigenvalue is therefore \((3+v)/4\) before dividing by \(4\), hence \((3+v)/8\) for \(A_I\). Thus
\begin{equation}
K=\frac{3+v}{8}\id
\end{equation}
is dual feasible, with
\begin{equation}
\Tr(K)=\frac{3+v}{4}.
\end{equation}
Therefore
\begin{equation}
D_{4,3}^{(\cQ)}(\brho_{\mathrm{tet}}^{(v)})=\frac{3+v}{4}.
\end{equation}

Averaging the three tetrahedral set-distinguishability values yields
\begin{equation}
\avgS_4^{(\cQ)}(\brho_{\mathrm{tet}}^{(v)})
=
\frac13\left(
\frac{1+v}{4}
+
\frac12+\frac{v}{2\sqrt3}
+
\frac{3+v}{4}
\right)
=
\frac12+\frac{1+1/\sqrt3}{6}v,
\end{equation}
and hence the stated expression for \(\Delta_4^{(\cQ)}(\brho_{\mathrm{tet}}^{(v)})\).

In both canonical families, the deviation is strictly positive for every \(v>0\), and vanishes only at the completely depolarized point \(v=0\).
\end{proof}

\subsection{Leakage families}

We now turn to a second natural deformation, in which an orthogonal branch is added that partially reveals the preparation label. Let \(\tau_1,\dots,\tau_n\) be perfectly distinguishable label states supported on a Hilbert-space summand orthogonal to that of the original ensemble \(\brho=(\rho_1,\dots,\rho_n)\). Define the corresponding leakage family by
\begin{equation}
\label{eq:leakage-family}
\Omega_i^{(\ell)}
=
(1-\ell)\,\rho_i\oplus 0
+
\ell\,0\oplus\tau_i,
\qquad
\ell\in[0,1].
\end{equation}

The key fact is that for orthogonal direct sums the relevant task-values decouple exactly.

\begin{lemma}[Direct-sum interpolation]
\label{lem:direct-sum}
Let \(\brho=(\rho_1,\dots,\rho_n)\) and \(\btau=(\tau_1,\dots,\tau_n)\) be ensembles supported on orthogonal Hilbert-space summands, and let \(\bOmega^{(\ell)}\) be defined by \eqref{eq:leakage-family}. Then for every \(m\in\{1,\dots,n-1\}\),
\begin{equation}
D_{n,m}^{(\cQ)}(\bOmega^{(\ell)})
=
(1-\ell)D_{n,m}^{(\cQ)}(\brho)
+
\ell D_{n,m}^{(\cQ)}(\btau).
\end{equation}
Consequently,
\begin{equation}
\avgD_n^{(\cQ)}(\bOmega^{(\ell)})
=
(1-\ell)\avgD_n^{(\cQ)}(\brho)
+
\ell\avgD_n^{(\cQ)}(\btau),
\end{equation}
and
\begin{equation}
\avgS_n^{(\cQ)}(\bOmega^{(\ell)})
=
(1-\ell)\avgS_n^{(\cQ)}(\brho)
+
\ell\avgS_n^{(\cQ)}(\btau).
\end{equation}
\end{lemma}

\begin{proof}
Because the states \(\Omega_i^{(\ell)}\) are block diagonal on orthogonal summands, only the block-diagonal part of any POVM contributes to the Born probabilities. Hence one may restrict without loss of generality to POVMs of the form
\begin{equation}
M_I = A_I \oplus B_I,
\qquad
A_I\succeq 0,\ \ B_I\succeq 0,
\qquad
\sum_I A_I=\id,\ \ \sum_I B_I=\id.
\end{equation}
For such a POVM,
\begin{align}
\sum_I\sum_{x\in I}\Tr(\Omega_x^{(\ell)} M_I)
&=
(1-\ell)\sum_I\sum_{x\in I}\Tr(\rho_x A_I)
+
\ell\sum_I\sum_{x\in I}\Tr(\tau_x B_I).
\end{align}
The two blocks involve disjoint optimization variables, subject to independent POVM normalization constraints on the two summands. Therefore the maximization splits exactly into two independent maximizations, yielding
\begin{equation}
D_{n,m}^{(\cQ)}(\bOmega^{(\ell)})
=
(1-\ell)D_{n,m}^{(\cQ)}(\brho)
+
\ell D_{n,m}^{(\cQ)}(\btau).
\end{equation}
The formulas for \(\avgD_n\) and \(\avgS_n\) follow immediately by averaging.
\end{proof}

For the perfectly revealing branch \(\btau\), every set-distinguishability task can be won perfectly:
\begin{equation}
D_{n,m}^{(\cQ)}(\btau)=1,
\qquad
m=1,\dots,n-1.
\end{equation}
Hence
\begin{equation}
\avgD_n^{(\cQ)}(\btau)=\avgS_n^{(\cQ)}(\btau)=1.
\end{equation}

\begin{theorem}[Leakage robustness]
\label{thm:leakage}
For the trine leakage family,
\begin{equation}
\Delta_3^{(\cQ)}(\bOmega_\triangle^{(\ell)})
=
(1-\ell)\,\frac{3\sqrt3-4}{12},
\end{equation}
and for the tetrahedral leakage family,
\begin{equation}
\Delta_4^{(\cQ)}(\bOmega_{\mathrm{tet}}^{(\ell)})
=
(1-\ell)\left(
\frac{\sqrt6}{6}
-\frac16
-\frac{1}{6\sqrt3}
\right).
\end{equation}
In particular, both canonical families violate the equality witness for every \(\ell<1\).
\end{theorem}

\begin{proof}
Apply Lemma~\ref{lem:direct-sum} to the trine and tetrahedral ensembles together with the perfectly revealing branch \(\btau\). Since \(\avgD_n^{(\cQ)}(\btau)=\avgS_n^{(\cQ)}(\btau)=1\), the revealing branch has zero deviation from the equality witness. Therefore
\begin{equation}
\Delta_n^{(\cQ)}(\bOmega^{(\ell)})
=
(1-\ell)\Delta_n^{(\cQ)}(\brho)+\ell\cdot 0
=
(1-\ell)\Delta_n^{(\cQ)}(\brho).
\end{equation}
Substituting the exact trine and tetrahedral values from Propositions~\ref{prop:trine-exact} and \ref{prop:tetra-exact} yields the stated formulas.
\end{proof}

The visibility and leakage theorems show that the positive-direction equality-witness violations are not fragile features of isolated symmetric ensembles. They persist throughout explicit one-parameter families all the way down to arbitrarily low positive visibility and up to arbitrarily large leakage short of complete disclosure. In this precise sense, the operational quantum description is epistemically incomplete already on simple and highly symmetric qubit fragments, and that incompleteness remains robust deep into noisy and high-leakage regimes.

\section{Sharpness via the Kochen--Specker \(\psi\)-epistemic model}

The Kochen--Specker \(\psi\)-epistemic model occupies a distinguished place in the present work. Historically, it is the canonical \(\psi\)-epistemic hidden-variable model for pure qubit states and projective measurements \cite{10.2307/24902153,Quanta22}. Here it acquires a sharper quantitative role. The quantum violations already certify epistemic incompleteness. What the Kochen--Specker model adds is sharpness: for the canonical trine and tetrahedral ensembles, it exactly saturates the hidden ontic excess implied by the equality violation. Thus, for these paradigmatic qubit examples, the equality witness is not merely a no-go statement against complete classical explanation; it identifies the exact amount of additional ontic communication power that ontological completion must expose.

\subsection{The Kochen--Specker qubit model}

The ontic state space is the Bloch sphere \(S^2\) equipped with the surface measure \(d\Omega\). A pure qubit state with Bloch vector \(\mathbf n\in S^2\) is represented by the epistemic density
\begin{equation}
\label{eq:KS-density}
\mu_{\mathbf n}(\boldsymbol\lambda)
=
\frac1\pi[\mathbf n\cdot\boldsymbol\lambda]_+,
\qquad
[x]_+:=\max\{x,0\}.
\end{equation}
Projective qubit measurements are represented in the standard way and reproduce the quantum Born probabilities \cite{Quanta22}. Throughout this section we evaluate the relevant ontic task-values directly by integration over \(S^2\).

\subsection{Pairwise discrimination}

\begin{lemma}[Pairwise discrimination in the Kochen--Specker model]
\label{lem:ks-pairwise}
Let \(\rho_{\mathbf n}\) and \(\rho_{\mathbf m}\) be pure qubit states whose Bloch vectors subtend an angle \(\alpha\in[0,\pi]\). Then
\begin{equation}
D_{2,1}^{(\Lambda_{\KS})}(\mu_{\mathbf n},\mu_{\mathbf m})
=
\frac12\left(1+\sin\frac{\alpha}{2}\right).
\end{equation}
In particular, this equals the quantum Helstrom optimum for the corresponding pure-state pair.
\end{lemma}

\begin{proof}
By rotational symmetry, place the two Bloch vectors in the equatorial plane at azimuths \(\pm\beta\) with \(\beta=\alpha/2\). Parametrize the ontic state as
\begin{equation}
\boldsymbol\lambda(\theta,\phi)
=
(\sin\theta\cos\phi,\sin\theta\sin\phi,\cos\theta),
\qquad
d\Omega=\sin\theta\,d\theta\,d\phi.
\end{equation}
The two epistemic densities become
\begin{equation}
\mu_\pm(\theta,\phi)=\frac1\pi\sin\theta\,[\cos(\phi\mp\beta)]_+.
\end{equation}
Hence
\begin{align}
D_{2,1}^{(\Lambda_{\KS})}
&=
\frac12\int_{S^2}d\Omega\,\max\{\mu_+,\mu_-\}\nonumber\\
&=
\frac12\cdot\frac1\pi
\left(\int_0^\pi \sin^2\theta\,d\theta\right)
\int_{-\pi}^{\pi}\max\{f_+(\phi),f_-(\phi)\}\,d\phi\nonumber\\
&=
\frac14\int_{-\pi}^{\pi}\max\{f_+(\phi),f_-(\phi)\}\,d\phi,
\end{align}
where \(f_\pm(\phi)=[\cos(\phi\mp\beta)]_+\) and \(\int_0^\pi\sin^2\theta\,d\theta=\pi/2\).

On the overlap interval,
\begin{equation}
\cos(\phi-\beta)-\cos(\phi+\beta)=2\sin\phi\,\sin\beta.
\end{equation}
Thus \(f_-(\phi)\ge f_+(\phi)\) for \(\phi\le 0\), while \(f_+(\phi)\ge f_-(\phi)\) for \(\phi\ge 0\). Therefore
\begin{equation}
\max\{f_+,f_-\}
=
\begin{cases}
\cos(\phi+\beta), & -\frac{\pi}{2}-\beta\le \phi\le 0,\\[0.4ex]
\cos(\phi-\beta), & 0\le \phi\le \frac{\pi}{2}+\beta,\\[0.4ex]
0, & \text{otherwise.}
\end{cases}
\end{equation}
Substituting,
\begin{align}
D_{2,1}^{(\Lambda_{\KS})}
&=
\frac14\left(
\int_{-\pi/2-\beta}^{0}\cos(\phi+\beta)\,d\phi
+
\int_0^{\pi/2+\beta}\cos(\phi-\beta)\,d\phi
\right)\nonumber\\
&=
\frac14\left[(1+\sin\beta)+(1+\sin\beta)\right]
=
\frac12(1+\sin\beta).
\end{align}
Since \(\beta=\alpha/2\), the lemma follows.
\end{proof}

This lemma is already important. It shows that for pure qubit pairs the Kochen--Specker model preserves the quantum pairwise discrimination power exactly. Any hidden excess revealed by the equality witness must therefore appear in the higher set-distinguishability quantities.

\subsection{Trine saturation}

\begin{theorem}[Trine saturation of the equality gap]
\label{thm:ks-trine}
Let \(\brho_\triangle=(\rho_1,\rho_2,\rho_3)\) be the trine ensemble and let \(\bmu_\triangle=(\mu_1,\mu_2,\mu_3)\) be its image in the Kochen--Specker model. Then
\begin{align}
\avgD_3^{(\Lambda_{\KS})}(\bmu_\triangle)
&=
\frac12\left(1+\frac{\sqrt3}{2}\right),\\
D_{3,1}^{(\Lambda_{\KS})}(\bmu_\triangle)
&=
\frac{\sqrt3}{2},\\
D_{3,2}^{(\Lambda_{\KS})}(\bmu_\triangle)
&=
1.
\end{align}
Consequently,
\begin{equation}
\avgS_3^{(\Lambda_{\KS})}(\bmu_\triangle)
=
\avgD_3^{(\Lambda_{\KS})}(\bmu_\triangle)
=
\frac12+\frac{\sqrt3}{4},
\end{equation}
and
\begin{equation}
\avgS_3^{(\Lambda_{\KS})}(\bmu_\triangle)-\avgS_3^{(\cQ)}(\brho_\triangle)
=
\avgD_3^{(\cQ)}(\brho_\triangle)-\avgS_3^{(\cQ)}(\brho_\triangle)
=
\frac{3\sqrt3-4}{12}.
\end{equation}
\end{theorem}

\begin{proof}
Every trine pair subtends the Bloch angle \(2\pi/3\), so Lemma~\ref{lem:ks-pairwise} gives
\begin{equation}
D_{2,1}^{(\Lambda_{\KS})}(\mu_i,\mu_j)
=
\frac12\left(1+\sin\frac{\pi}{3}\right)
=
\frac12\left(1+\frac{\sqrt3}{2}\right),
\end{equation}
and averaging over the three pairs yields the first claim.

For \(D_{3,1}^{(\Lambda_{\KS})}\), choose azimuths
\begin{equation}
\varphi_1=0,\qquad \varphi_2=\frac{2\pi}{3},\qquad \varphi_3=-\frac{2\pi}{3},
\end{equation}
and write
\begin{equation}
\mu_i(\theta,\phi)=\frac1\pi\sin\theta\,f_i(\phi),
\qquad
f_i(\phi)=[\cos(\phi-\varphi_i)]_+.
\end{equation}
Then
\begin{equation}
D_{3,1}^{(\Lambda_{\KS})}
=
\frac13\int_{S^2}d\Omega\,\max\{\mu_1,\mu_2,\mu_3\}
=
\frac16\int_{-\pi}^{\pi}\max\{f_1,f_2,f_3\}\,d\phi.
\end{equation}
On the sector \([-\pi/3,\pi/3]\), \(f_1(\phi)=\cos\phi\) dominates both \(f_2\) and \(f_3\). By \(2\pi/3\) rotational symmetry,
\begin{equation}
\int_{-\pi}^{\pi}\max\{f_1,f_2,f_3\}\,d\phi
=
3\int_{-\pi/3}^{\pi/3}\cos\phi\,d\phi
=
3\sqrt3.
\end{equation}
Therefore
\begin{equation}
D_{3,1}^{(\Lambda_{\KS})}=\frac{\sqrt3}{2}.
\end{equation}

For \(D_{3,2}^{(\Lambda_{\KS})}\), at every point of the circle at least one of the three functions \(f_i\) vanishes, because the three positive-support semicircles have empty triple intersection. Hence
\begin{equation}
\max\{f_1+f_2,\ f_1+f_3,\ f_2+f_3\}=f_1+f_2+f_3
\end{equation}
pointwise. Therefore
\begin{align}
D_{3,2}^{(\Lambda_{\KS})}
&=
\frac13\int_{S^2}d\Omega\,\max\{\mu_1+\mu_2,\mu_1+\mu_3,\mu_2+\mu_3\}\nonumber\\
&=
\frac16\int_{-\pi}^{\pi}(f_1+f_2+f_3)\,d\phi.
\end{align}
Each \(f_i\) integrates to \(2\), so the total integral is \(6\), and thus \(D_{3,2}^{(\Lambda_{\KS})}=1\).

Averaging \(D_{3,1}^{(\Lambda_{\KS})}\) and \(D_{3,2}^{(\Lambda_{\KS})}\) yields \(\avgS_3^{(\Lambda_{\KS})}\). The equality \(\avgS_3^{(\Lambda_{\KS})}=\avgD_3^{(\Lambda_{\KS})}\) also follows from Theorem~\ref{thm:ontic-equality}. Finally, substituting the exact quantum values from Proposition~\ref{prop:trine-exact} gives the claimed tightness relation.
\end{proof}

The trine case already shows the essential point. The equality violation does not merely imply that some hidden excess ontic communication power must exist; it determines an exact amount, and the Kochen--Specker model realizes precisely that amount.

\subsection{Tetrahedral saturation}

\begin{theorem}[Tetrahedral saturation of the equality gap]
\label{thm:ks-tetra}
Let \(\brho_{\mathrm{tet}}=(\rho_1,\rho_2,\rho_3,\rho_4)\) be the tetrahedral ensemble and let \(\bmu_{\mathrm{tet}}\) be its image in the Kochen--Specker model. Then
\begin{align}
\avgD_4^{(\Lambda_{\KS})}(\bmu_{\mathrm{tet}})
&=
\frac12\left(1+\sqrt{\frac23}\right),\\
D_{4,1}^{(\Lambda_{\KS})}(\bmu_{\mathrm{tet}})
&=
\frac{\sqrt6}{\pi}\arctan(\sqrt2),\\
D_{4,3}^{(\Lambda_{\KS})}(\bmu_{\mathrm{tet}})
&=
1,\\
D_{4,2}^{(\Lambda_{\KS})}(\bmu_{\mathrm{tet}})
&=
\frac12+\frac32\sqrt{\frac23}-\frac{\sqrt6}{\pi}\arctan(\sqrt2).
\end{align}
Consequently,
\begin{equation}
\avgS_4^{(\Lambda_{\KS})}(\bmu_{\mathrm{tet}})
=
\avgD_4^{(\Lambda_{\KS})}(\bmu_{\mathrm{tet}})
=
\frac12\left(1+\sqrt{\frac23}\right),
\end{equation}
and
\begin{equation}
\avgS_4^{(\Lambda_{\KS})}(\bmu_{\mathrm{tet}})-\avgS_4^{(\cQ)}(\brho_{\mathrm{tet}})
=
\avgD_4^{(\cQ)}(\brho_{\mathrm{tet}})-\avgS_4^{(\cQ)}(\brho_{\mathrm{tet}})
=
\Delta_4^{(\cQ)}(\brho_{\mathrm{tet}}).
\end{equation}
\end{theorem}

\begin{proof}
Every tetrahedral pair has overlap squared \(1/3\), so Lemma~\ref{lem:ks-pairwise} immediately yields
\begin{equation}
\avgD_4^{(\Lambda_{\KS})}(\bmu_{\mathrm{tet}})
=
\frac12\left(1+\sqrt{\frac23}\right).
\end{equation}

To compute \(D_{4,1}^{(\Lambda_{\KS})}\), choose coordinates so that
\begin{equation}
\mathbf t_1=(0,0,1),
\end{equation}
while the remaining three Bloch vectors lie at polar angle \(\alpha\) with
\begin{equation}
\cos\alpha=-\frac13,\qquad
\sin\alpha=\frac{2\sqrt2}{3},
\end{equation}
and azimuths \(0,2\pi/3,4\pi/3\). By symmetry,
\begin{equation}
D_{4,1}^{(\Lambda_{\KS})}
=
\int_{V_1}\mu_1(\boldsymbol\lambda)\,d\Omega,
\end{equation}
where \(V_1\) is the spherical Voronoi cell on which \(\mu_1\) dominates. On the fundamental angular sector \(0\le\phi\le\pi/3\), the cell boundary is determined by \(\mu_1=\mu_2\), which reduces to
\begin{equation}
\tan\theta=\frac{\sqrt2}{\cos\phi}.
\end{equation}
Since the full cell has six congruent sectors,
\begin{align}
D_{4,1}^{(\Lambda_{\KS})}
&=
\frac{6}{\pi}
\int_0^{\pi/3}\int_0^{\theta_b(\phi)}
\cos\theta\,\sin\theta\,d\theta\,d\phi,\qquad
\theta_b(\phi)=\arctan\!\left(\frac{\sqrt2}{\cos\phi}\right)\nonumber\\
&=
\frac{3}{\pi}\int_0^{\pi/3}\sin^2\theta_b(\phi)\,d\phi.
\end{align}
Using
\begin{equation}
\sin^2\theta_b=\frac{2}{2+\cos^2\phi}
\end{equation}
and the substitution \(u=\tan\phi\), one finds
\begin{equation}
\int_0^{\pi/3}\frac{2}{2+\cos^2\phi}\,d\phi
=
\sqrt{\frac23}\,\arctan(\sqrt2),
\end{equation}
hence
\begin{equation}
D_{4,1}^{(\Lambda_{\KS})}
=
\frac{\sqrt6}{\pi}\arctan(\sqrt2).
\end{equation}

For \(D_{4,3}^{(\Lambda_{\KS})}\), note that the tetrahedral Bloch vectors sum to zero:
\begin{equation}
\sum_{i=1}^4 \mathbf t_i=0.
\end{equation}
Therefore, for every ontic direction \(\boldsymbol\lambda\), not all four inner products \(\mathbf t_i\cdot\boldsymbol\lambda\) can be strictly positive. Hence at least one of the densities \(\mu_i(\boldsymbol\lambda)\) vanishes, so the best three-subset sum is just the total sum:
\begin{equation}
\max_{|I|=3}\sum_{i\in I}\mu_i(\boldsymbol\lambda)
=
\sum_{i=1}^4\mu_i(\boldsymbol\lambda).
\end{equation}
Integrating and using normalization of each \(\mu_i\) gives \(D_{4,3}^{(\Lambda_{\KS})}=1\).

To determine \(D_{4,2}^{(\Lambda_{\KS})}\), invoke Theorem~\ref{thm:ontic-equality}. Since
\begin{equation}
\avgS_4^{(\Lambda_{\KS})}
=
\frac13\left(D_{4,1}^{(\Lambda_{\KS})}+D_{4,2}^{(\Lambda_{\KS})}+D_{4,3}^{(\Lambda_{\KS})}\right)
\end{equation}
and
\begin{equation}
\avgS_4^{(\Lambda_{\KS})}
=
\avgD_4^{(\Lambda_{\KS})}
=
\frac12\left(1+\sqrt{\frac23}\right),
\end{equation}
substituting the already determined values of \(D_{4,1}^{(\Lambda_{\KS})}\) and \(D_{4,3}^{(\Lambda_{\KS})}\) and solving for \(D_{4,2}^{(\Lambda_{\KS})}\) yields the claimed expression. The equality \(\avgS_4^{(\Lambda_{\KS})}=\avgD_4^{(\Lambda_{\KS})}\) is then immediate, and the final tightness relation follows by substituting the exact quantum values from Proposition~\ref{prop:tetra-exact}.
\end{proof}

\begin{remark}[Why the saturation result matters]
For the trine ensemble, the hidden ontic excess forced by the equality violation is concentrated entirely in the full three-label task \(D_{3,1}\). For the tetrahedral ensemble, the hidden ontic excess is distributed between \(D_{4,1}\) and \(D_{4,2}\). In both cases, however, the total hidden excess is exactly equal to the quantum equality gap. The equality witness is therefore not merely qualitative: for these canonical qubit examples it is an exact quantitative accounting identity for hidden ontic communication power.
\end{remark}

These saturation results complete the foundational picture for the canonical qubit examples. The equality gap is not merely a sign of epistemic incompleteness; it is an exact amount of hidden ontic task-power. We now turn to the broader numerical problem of determining how large the quantum deviation can be in general.

\section{Characterising and bounding the incompleteness of quantum theory}
\label{CABIOQT}

The preceding sections establish that quantum theory violates the equality witness, that the violation can occur in both directions, that the canonical trine and tetrahedral qubit ensembles furnish exact positive-direction examples, that these violations persist throughout explicit visibility and leakage families, and that the Kochen--Specker model saturates the hidden ontic excess in the canonical qubit cases. What remains open is the global optimization problem: how large can the quantum deviation from the equality witness be when one optimizes over \emph{all} finite preparation sets and \emph{all} finite-dimensional realizations?

The purpose of the present section is to formulate that optimization problem in a way that can be attacked systematically, and to record what the resulting semidefinite-programming hierarchy and see-saw methods certify in the low-cardinality cases relevant to the manuscript. These numerical methods are not needed to establish the witness itself or its sharpness for the canonical qubit ensembles. Rather, they serve a complementary role: they bound the full quantum region compatible with the equality witness, certify the optimality of the canonical positive-direction points at small \(n\), bound the opposite-sign extremum, and provide evidence for how the maximal deviation behaves as the number of preparations increases.

The hierarchy architecture used below follows the moment-matrix approach originating in the Navascu\'es--Pironio--Ac\'in hierarchy for nonlocal quantum correlations and its contextuality and prepare-and-measure adaptations \cite{Navascues2007,navascues2008convergent,Chaturvedi2021characterising,PRXQuantum.2.020334,Tavakoli2022informationally}. In parallel, the use of task-value benchmarks and auxiliary operators is aligned with informationally restricted prepare-and-measure methods \cite{Tavakoli2020informationally,Tavakoli2022informationally,PauwelsPironioTavakoli2025}.

For positive deviations, one considers the constrained optimization problem
\begin{equation}
\label{eq:opt-positive}
\bar{D}^{(\cQ)}_{n}(p)
=
\max_{\bm{\rho}}\ \avgD_n^{(\cQ)}(\bm{\rho})
\qquad
\text{s.t.}\qquad
\avgS_n^{(\cQ)}(\bm{\rho})\le p,
\end{equation}
and for negative deviations the dual constrained problem
\begin{equation}
\label{eq:opt-negative}
\bar{S}^{(\cQ)}_{n}(p)
=
\max_{\bm{\rho}}\ \avgS_n^{(\cQ)}(\bm{\rho})
\qquad
\text{s.t.}\qquad
\avgD_n^{(\cQ)}(\bm{\rho})\le p.
\end{equation}
Here \(\bm{\rho}\equiv\{\rho_x\in\mathcal{B}_+(\mathcal{H})\,|\,\Tr\rho_x=1\}_{x=1}^n\) denotes an arbitrary set of \(n\) quantum preparations of unrestricted Hilbert-space dimension. The associated maximal positive and negative deviations are
\begin{align}
\Delta^{\max}_{+}(n)
&:=
\sup_{\bm{\rho}}\Bigl\{\avgD_n^{(\cQ)}(\bm{\rho})-\avgS_n^{(\cQ)}(\bm{\rho})\Bigr\},\\
\Delta^{\max}_{-}(n)
&:=
\sup_{\bm{\rho}}\Bigl\{\avgS_n^{(\cQ)}(\bm{\rho})-\avgD_n^{(\cQ)}(\bm{\rho})\Bigr\},\\
\Delta^{\max}_{\mathrm{abs}}
&:=
\sup_{n\ge 2}\max\!\Bigl\{\Delta^{\max}_{+}(n),\Delta^{\max}_{-}(n)\Bigr\}.
\end{align}
The last quantity is the natural candidate for the absolute maximal epistemic incompleteness of quantum theory.

We now present the semidefinite-programming machinery in full. We treat first the positive-side optimization family \eqref{eq:opt-positive} in detail, and then describe the corresponding modifications for the negative-side family \eqref{eq:opt-negative}.

\subsection{Average pairwise distinguishability given average set-distinguishability}

We start from the optimization problem \eqref{eq:opt-positive}. Recall that
\begin{equation}
\avgS_n^{(\cQ)}(\bm{\rho})
=
\frac{1}{n-1}\sum_{m=1}^{n-1} D_{n,m}^{(\cQ)}(\bm{\rho}),
\end{equation}
where for any \(m\in[n-1]\),
\begin{equation}
\label{eq:nmSD}
\begin{split}
D_{n,m}^{(\cQ)}(\bm{\rho})
=
\frac1n \max_{\{M^{(m)}_I\}}
& \sum_{\substack{I\subset[n]\\ |I|=m}}
\sum_{x\in I}\Tr(\rho_x M^{(m)}_{I})\\
\text{s.t.}\qquad
& M^{(m)}_{I}\succeq 0,\qquad \forall\, I\subset[n],\ |I|=m,\\
& \sum_{\substack{I\subset[n]\\ |I|=m}} M^{(m)}_I=\id.
\end{split}
\end{equation}

Following the architecture of informationally restricted prepare-and-measure SDPs \cite{Tavakoli2022informationally}, we introduce \(n-1\) auxiliary operators \(\bm{\Theta}\equiv\{\Theta_m\succeq 0\}_{m=1}^{n-1}\) satisfying
\begin{equation}
\label{eq:auxSDPConstraint}
\Theta_m \succeq \sum_{x\in I}\rho_x,
\qquad
\forall\, m\in[n-1],\ \forall\, I\subset[n],\ |I|=m.
\end{equation}
These operators arise from the dual of the optimization problem \eqref{eq:nmSD}. They allow us to upper-bound the average set-distinguishability in the form
\begin{align}
\avgS_n^{(\cQ)}(\bm{\rho})
&=
\frac{1}{n(n-1)}\sum_{m=1}^{n-1}
\max_{\{M_I^{(m)}\}}
\sum_{\substack{I\subset[n]\\ |I|=m}}
\Tr\!\left(\Bigl(\sum_{x\in I}\rho_x\Bigr)M_I^{(m)}\right)\nonumber\\
&\le
\frac{1}{n(n-1)}\sum_{m=1}^{n-1}
\max_{\{M_I^{(m)}\}}
\sum_{\substack{I\subset[n]\\ |I|=m}}
\Tr(\Theta_m M_I^{(m)})\nonumber\\
&=
\frac{1}{n(n-1)}\sum_{m=1}^{n-1}\Tr(\Theta_m),
\end{align}
where in the last step we used \(\sum_{I:|I|=m}M_I^{(m)}=\id\) for each \(m\). Hence the benchmark constraint \(\avgS_n^{(\cQ)}(\bm{\rho})\le p\) can be enforced through the tracial condition
\begin{equation}
\label{eq:tracialConstraint}
\frac{1}{n(n-1)}\sum_{m=1}^{n-1}\Tr(\Theta_m)\le p
\end{equation}
together with the semidefinite dominance constraints \eqref{eq:auxSDPConstraint}.

The introduction of the auxiliary operators \(\bm{\Theta}\) is the key ingredient enabling both dimension-dependent lower bounds and dimension-independent upper bounds on the constrained optimum \(\bar{D}^{(\cQ)}_n(p)\).

\subsubsection{Alternating semidefinite programs for dimension-dependent lower bounds}

With the auxiliary operators in place, the positive-side optimization may be rewritten as
\begin{equation}
\label{eq:APWDGivenASDSeeSaw}
\begin{split}
\bar{D}^{(\cQ)}_{n}(p)=
\max_{\bm{\rho},\,\{M^{(i,j)}_k\},\,\bm{\Theta}}
&\ \frac{1}{2\binom{n}{2}}
\sum_{i<j}\sum_{x\in\{i,j\}}
\Tr(\rho_x M^{(i,j)}_{k=x})\\
\text{s.t.}\qquad
& \Theta_m \succeq \sum_{x\in I}\rho_x,
\qquad \forall\, m\in[n-1],\ \forall\, I\subset[n],\ |I|=m,\\
& \frac{1}{n(n-1)}\sum_{m=1}^{n-1}\Tr(\Theta_m)\le p,\\
& \rho_x\succeq 0,\qquad \Tr\rho_x=1,\qquad \forall\, x\in[n],\\
& M^{(i,j)}_{k}\succeq 0,\qquad \forall\, k\in\{i,j\},\ \forall\, i<j,\\
& \sum_{k\in\{i,j\}}M^{(i,j)}_k=\id,\qquad \forall\, i<j.
\end{split}
\end{equation}
For fixed measurements \(\{M^{(i,j)}_k\}\), this is a semidefinite program in the states \(\bm{\rho}\) and auxiliary operators \(\bm{\Theta}\). For fixed states and auxiliary operators, it becomes a semidefinite program in the pairwise Helstrom measurements \(\{M^{(i,j)}_k\}\). Alternating these two semidefinite programs therefore yields a dimension-dependent feasible lower bound \(\bar{D}^{(\cQ_{\mathrm{LB}})}_n(p)\le \bar{D}^{(\cQ)}_n(p)\). Increasing the allowed Hilbert-space dimension can only improve the best lower bound found.

\subsubsection{Hierarchy of SDP relaxations for dimension-independent upper bounds}

The optimization problem \eqref{eq:APWDGivenASDSeeSaw} is difficult primarily because the Hilbert-space dimension is unrestricted. To obtain dimension-independent upper bounds, we follow the moment-matrix strategy used in nonlocal, contextual, and informationally restricted prepare-and-measure scenarios \cite{Navascues2007,navascues2008convergent,Chaturvedi2021characterising,PRXQuantum.2.020334,Tavakoli2022informationally}.

For a positive semidefinite operator \(\tau\succeq 0\) and an ordered list of operators \(\mathcal{O}=(O_1,O_2,\dots)\), define the moment matrix
\begin{equation}
\Gamma^{\mathcal{O}}_{\tau}
\quad\text{with entries}\quad
(\Gamma^{\mathcal{O}}_{\tau})_{i,j}:=\Tr(\tau\,O_i^\dagger O_j).
\end{equation}
Positivity of \(\tau\) implies \(\Gamma^{\mathcal{O}}_{\tau}\succeq 0\).

In our setting we use \(n\) moment matrices \(\{\Gamma^{\mathcal{O}}_{\rho_x}\}_{x=1}^n\) hinged on the physical states and \(n-1\) moment matrices \(\{\Gamma^{\mathcal{O}}_{\Theta_m}\}_{m=1}^{n-1}\) hinged on the auxiliary operators. The semidefinite dominance constraints \eqref{eq:auxSDPConstraint} lift directly to
\begin{equation}
\label{eq:auxSDPConstraintGamma}
\Gamma^{\mathcal{O}}_{\Theta_m}
\succeq
\sum_{x\in I}\Gamma^{\mathcal{O}}_{\rho_x},
\qquad
\forall\, m\in[n-1],\ \forall\, I\subset[n],\ |I|=m.
\end{equation}

At level \(1\), let \(\mathcal{O}_1\) contain the identity and the one-letter monomials built from the binary Helstrom projectors \(\{M^{(i,j)}_{k=i}\}_{i<j}\):
\begin{equation}
\mathcal{O}_1:=\bigl(\id,\{M^{(i,j)}_{k=i}\}_{i<j\in[n]}\bigr).
\end{equation}
Then the observed probabilities appear directly as moment-matrix entries,
\begin{equation}
(\Gamma^{\mathcal{O}_1}_{\rho_x})_{\id,M^{(i,j)}_{k=i}}
=
\Tr(\rho_x M^{(i,j)}_{k=i}),
\end{equation}
while the tracial benchmark becomes
\begin{equation}
\frac{1}{n(n-1)}\sum_{m=1}^{n-1}(\Gamma^{\mathcal{O}_1}_{\Theta_m})_{\id,\id}\le p.
\end{equation}
Because there is no dimension restriction, Naimark dilation allows us to take the certifying measurements projective without loss of generality. Hence
\begin{equation}
(M^{(i,j)}_{k=i})^\dagger M^{(i,j)}_{k=i}=M^{(i,j)}_{k=i},
\end{equation}
which translates into linear constraints such as
\begin{equation}
(\Gamma^{\mathcal{O}_1}_{\tau})_{M^{(i,j)}_{k=i},M^{(i,j)}_{k=i}}
=
(\Gamma^{\mathcal{O}_1}_{\tau})_{\id,M^{(i,j)}_{k=i}}
\end{equation}
for every \(\tau\in\{\rho_x\}_{x=1}^n\cup\{\Theta_m\}_{m=1}^{n-1}\).

The resulting first-level relaxation is
\begin{equation}
\label{eq:APWDGivenASDQ1}
\begin{split}
\bar{D}^{(\cQ_1)}_{n}(p)=
\max_{\{\Gamma_{\rho_x}^{\mathcal{O}_1}\},\,\{\Gamma_{\Theta_m}^{\mathcal{O}_1}\}}
&\ \frac{1}{2}\left(
1+\frac{1}{\binom{n}{2}}
\sum_{i<j}
\Bigl[
(\Gamma_{\rho_i}^{\mathcal{O}_1})_{\id,M^{(i,j)}_{k=i}}
-
(\Gamma_{\rho_j}^{\mathcal{O}_1})_{\id,M^{(i,j)}_{k=i}}
\Bigr]
\right)\\
\text{s.t.}\qquad
& \Gamma_{\Theta_m}^{\mathcal{O}_1}
\succeq
\sum_{x\in I}\Gamma_{\rho_x}^{\mathcal{O}_1},
\qquad \forall\, m,\ \forall\, I,\ |I|=m,\\
& \frac{1}{n(n-1)}\sum_{m=1}^{n-1}
(\Gamma_{\Theta_m}^{\mathcal{O}_1})_{\id,\id}\le p,\\
& \Gamma_{\rho_x}^{\mathcal{O}_1}\succeq 0,\qquad
(\Gamma_{\rho_x}^{\mathcal{O}_1})_{\id,\id}=1,\qquad \forall\, x,\\
& \Gamma_{\Theta_m}^{\mathcal{O}_1}\succeq 0,\qquad \forall\, m,\\
& (\Gamma_{\tau}^{\mathcal{O}_1})_{M^{(i,j)}_{k=i},M^{(i,j)}_{k=i}}
=
(\Gamma_{\tau}^{\mathcal{O}_1})_{\id,M^{(i,j)}_{k=i}},
\qquad \forall\, i<j,\ \forall\, \tau.
\end{split}
\end{equation}
This is a relaxation of the true quantum optimization, and therefore
\begin{equation}
\bar{D}^{(\cQ_1)}_{n}(p)\ge \bar{D}^{(\cQ)}_n(p)\ge \bar{D}^{(\cQ_{\mathrm{LB}})}_n(p).
\end{equation}

Higher levels are obtained by taking \(\mathcal{O}_{\mathcal{L}}\) to be the list of monomials of elements of \(\mathcal{O}_1\) of length at most \(\mathcal{L}\). This yields a hierarchy of successively tightening dimension-independent upper bounds
\begin{equation}
\bar{D}^{(\cQ_{\mathcal{L}})}_n(p)
\ge
\bar{D}^{(\cQ_{\mathcal{L}+1})}_n(p)
\ge
\bar{D}^{(\cQ)}_n(p)
\ge
\bar{D}^{(\cQ_{\mathrm{LB}})}_n(p)
\qquad
\forall\, \mathcal{L}\in\mathbb{N}.
\end{equation}

Finally, for a fixed \(n\), a slight modification of the same relaxation directly upper-bounds the maximal positive deviation itself. One replaces the objective function in \eqref{eq:APWDGivenASDQ1} by
\begin{equation}
\Delta^{(\cQ_{\mathcal{L}})}_n
=
\max\Bigl\{
\bar{D}^{(\cQ_{\mathcal{L}})}_n(p)-p
\Bigr\},
\end{equation}
now treating \(p\ge 0\) as an optimization variable subject to the same tracial benchmark constraint. This gives a certified dimension-independent upper bound on the maximal positive deviation for the \(n\)-preparation equality.

\subsection{Average set-distinguishability given average pairwise distinguishability}

We now consider the dual optimization family \eqref{eq:opt-negative}. The corresponding machinery is analogous, so we record only the key ingredient enabling the benchmark constraint.

Recall that
\begin{equation}
\avgD_n^{(\cQ)}(\bm{\rho})
=
\frac{1}{\binom{n}{2}}
\sum_{i<j} D_{2,1}^{(\cQ)}(\rho_i,\rho_j),
\end{equation}
where
\begin{equation}
\label{eq:ijPD}
\begin{split}
D_{2,1}^{(\cQ)}(\rho_i,\rho_j)
=
\frac12\max_{\{M^{(i,j)}_k\}}
& \sum_{x\in\{i,j\}}\Tr(\rho_x M^{(i,j)}_{k=x})\\
\text{s.t.}\qquad
& M^{(i,j)}_k\succeq 0,\qquad \forall\, k\in\{i,j\},\\
& \sum_{k\in\{i,j\}}M^{(i,j)}_k=\id.
\end{split}
\end{equation}
To impose the benchmark \(\avgD_n^{(\cQ)}(\bm{\rho})\le p\), introduce \(\binom{n}{2}\) auxiliary operators
\begin{equation}
\bm{\Theta}\equiv\{\Theta_{ij}\succeq 0\}_{i<j}
\end{equation}
with the property
\begin{equation}
\label{eq:auxSDPConstraint1}
\Theta_{ij}\succeq \rho_i,\qquad \Theta_{ij}\succeq \rho_j,\qquad \forall\, i<j.
\end{equation}
Then
\begin{align}
\avgD_n^{(\cQ)}(\bm{\rho})
&=
\frac{1}{2\binom{n}{2}}
\sum_{i<j}
\max_{\{M^{(i,j)}_k\}}
\sum_{x\in\{i,j\}}\Tr(\rho_x M^{(i,j)}_{k=x})\nonumber\\
&\le
\frac{1}{2\binom{n}{2}}
\sum_{i<j}
\max_{\{M^{(i,j)}_k\}}
\sum_{x\in\{i,j\}}\Tr(\Theta_{ij} M^{(i,j)}_{k=x})\nonumber\\
&=
\frac{1}{2\binom{n}{2}}
\sum_{i<j}\Tr(\Theta_{ij}).
\end{align}
Hence the benchmark \(\avgD_n^{(\cQ)}(\bm{\rho})\le p\) can be enforced through the tracial constraint
\begin{equation}
\label{eq:tracialConstraint1}
\frac{1}{2\binom{n}{2}}\sum_{i<j}\Tr(\Theta_{ij})\le p
\end{equation}
together with the semidefinite constraints \eqref{eq:auxSDPConstraint1}.

Combining \eqref{eq:auxSDPConstraint1} and \eqref{eq:tracialConstraint1} with the averaged set-distinguishability objective yields the negative-side analogue of the see-saw formulation \eqref{eq:APWDGivenASDSeeSaw}. For fixed set-distinguishability measurements, the problem is a semidefinite program in the states and auxiliary operators; for fixed states and auxiliary operators, it decomposes into semidefinite programs over the certifying measurement families. In parallel, the corresponding moment-matrix hierarchy is obtained by introducing moment matrices hinged on the states and on the auxiliary operators \(\Theta_{ij}\), together with operator lists containing the one-letter monomials of the relevant set-distinguishability measurement effects. This yields dimension-dependent lower bounds \(\bar{S}^{(\cQ_{\mathrm{LB}})}_n(p)\) and a tightening sequence of dimension-independent upper bounds \(\bar{S}^{(\cQ_{\mathcal{L}})}_n(p)\).

\subsection{Certified extremal frontiers for \(n=3\) and \(n=4\)}

We now state precisely what the hierarchy and see-saw methods certify in the low-cardinality cases used in the manuscript.

\paragraph*{Three preparations: positive boundary.}
For \(n=3\), the positive-side optimization does more than certify the trine point itself. The third level of the hierarchy matches the see-saw lower bounds along the entire optimal positive boundary in the \((\avgS_3,\avgD_3)\) plane. This boundary is generated explicitly by the trine visibility and leakage families.

Define
\begin{equation}
\bar{D}^{\max}_{3}(p)
:=
\sup_{\bm{\rho}}
\Bigl\{
\avgD_3^{(\cQ)}(\bm{\rho})
\,:\,
\avgS_3^{(\cQ)}(\bm{\rho})\le p
\Bigr\}.
\end{equation}
Then the certified optimum is
\begin{equation}
\bar{D}^{\max}_{3}(p)
=
\frac12+\frac{3\sqrt3}{4}\left(p-\frac12\right),
\qquad
\frac12\le p\le \frac56,
\end{equation}
and this branch is attained by the visibility-deformed trine family \(\brho_\triangle^{(v)}\), with
\begin{equation}
v=3p-\frac32.
\end{equation}
For the remaining benchmark range,
\begin{equation}
\bar{D}^{\max}_{3}(p)
=
1-\frac32(2-\sqrt3)(1-p),
\qquad
\frac56\le p\le 1,
\end{equation}
and this branch is attained by the leakage trine family \(\bOmega_\triangle^{(\ell)}\), with
\begin{equation}
\ell=6p-5.
\end{equation}
In particular, at the trine benchmark
\begin{equation}
p=\avgS_3^{(\cQ)}(\brho_\triangle)=\frac56,
\end{equation}
one recovers the exact maximal positive deviation
\begin{equation}
\Delta^{\max}_{+}(3)
=
\Delta_3^{(\cQ)}(\brho_\triangle)
=
\frac{3\sqrt3-4}{12}.
\end{equation}
Thus the trine ensemble is the exact positive-direction maximizer of the three-preparation equality, and the explicit visibility and leakage families describe the full optimal positive boundary for \(n=3\).

\paragraph*{Three preparations: negative side.}
For the inverse optimization problem at \(n=3\), the third level of the hierarchy yields an upper bound that coincides, up to machine precision, with the best lower bound found by the see-saw method. The optimum is attained by a triplet of mixed qutrit states \(\bm{\rho}_{\mathrm{neg}}\), with
\begin{equation}
\avgS_3^{(\cQ)}(\bm{\rho}_{\mathrm{neg}})
-
\avgD_3^{(\cQ)}(\bm{\rho}_{\mathrm{neg}})
\approx 0.0277
\end{equation}
at the benchmark value
\begin{equation}
p=\avgD_3^{(\cQ)}(\bm{\rho}_{\mathrm{neg}})\approx 0.8214.
\end{equation}
Hence the witness is genuinely bidirectional already at \(n=3\). At the same time, the maximal absolute deviation for the three-preparation equality is the positive trine value:
\begin{equation}
\max\!\Bigl\{\Delta^{\max}_{+}(3),\Delta^{\max}_{-}(3)\Bigr\}
=
\Delta^{\max}_{+}(3)
=
\frac{3\sqrt3-4}{12}.
\end{equation}

\paragraph*{Four preparations: positive side.}
For the four-preparation equality, the second level of the positive-side hierarchy numerically certifies the extremality of the tetrahedral ensemble by matching its exact analytical value with the see-saw lower bound up to machine precision. Concretely, at
\begin{equation}
p=\avgS_4^{(\cQ)}(\brho_{\mathrm{tet}})
=
\frac23+\frac{1}{6\sqrt3},
\end{equation}
the certified optimum is
\begin{equation}
\Delta^{\max}_{+}(4)
=
\Delta_4^{(\cQ)}(\brho_{\mathrm{tet}})
=
\frac12\left(1+\sqrt{\frac23}\right)-\frac23-\frac{1}{6\sqrt3}
\approx 0.14535658
\end{equation}
up to machine precision. Thus the tetrahedral ensemble is the numerically certified positive-direction maximizer of the \(n=4\) equality.

Taken together, these results establish the low-cardinality extremal picture used throughout the manuscript: the trine ensemble is the exact positive-direction maximizer for \(n=3\), the tetrahedral ensemble is the numerically certified positive-direction maximizer for \(n=4\), and the negative-direction extremum for \(n=3\) is realized by a mixed qutrit ensemble.

\subsection{Increasing maximal positive deviation}

The same methods also provide evidence that the maximal positive deviation increases with the number of preparations in the cases studied.

\begin{figure}[t]
\centering
\includegraphics[width=0.65\linewidth]{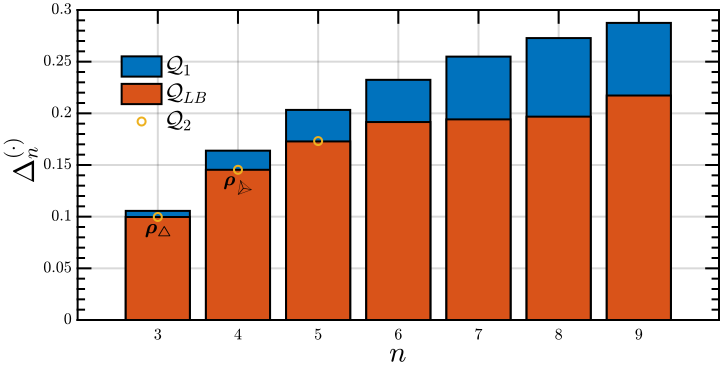}
\caption{Increasing maximal positive quantum deviation from the equality witness. The values shown are the exact trine and tetrahedral deviations for \(n=3\) and \(n=4\), together with the numerically certified value displayed for \(n=5\). In particular, the \(n=3\) maximum is attained exactly by the trine ensemble, while the \(n=4\) maximum is numerically certified at the tetrahedral ensemble by hierarchy/see-saw matching. The observed trend suggests that epistemic incompleteness becomes stronger with larger preparation sets and motivates the open question of the absolute maximal epistemic incompleteness of quantum theory.}
\label{fig:increasing}
\end{figure}

This evidence points to the sharper global problem of determining
\begin{equation}
\Delta^{\max}_{\mathrm{abs}}
=
\sup_{n\ge 2}\max\!\Bigl\{\Delta^{\max}_{+}(n),\Delta^{\max}_{-}(n)\Bigr\},
\end{equation}
namely the absolute maximal epistemic incompleteness of quantum theory, obtained by maximizing the equality-witness deviation over all finite preparation sets and over all \(n\).

\section{Hierarchy of classicality notions and unifying implications}

The equality witness already stands on its own as a completeness test, a resource witness, and, through its nonzero violations, a witness of coherence and measurement incompatibility. What remains is to place the underlying notion of epistemic completeness within the broader landscape of classicality principles. The point of the present section is not merely classificatory. Rather, it is to make precise the unifying role of epistemic completeness: because it requires exact preservation of the full family of empirical communication-task quantities of a finite preparation set, it sits above several more familiar notions that preserve only selected operational features. What drives the hierarchy is precisely this difference between preserving the full family and preserving only selected operational shadows of it. Consequently, whenever one of those weaker notions is ruled out operationally, epistemic completeness is ruled out as well.

At the level of unconditional preparation-side implications, the logical core is
\begin{equation}
\label{eq:hierarchy-core}
\text{epistemic completeness}
\Longrightarrow
\text{bounded ontological distinctness on preparations}
\Longrightarrow
\text{preparation noncontextuality}.
\end{equation}
When one further restricts to the standard pure-state / projective-measurement setting in quantum theory, familiar conditional routes to maximally \(\psi\)-epistemic structure and Kochen--Specker noncontextuality enter as well, as discussed below. The purpose of Eq.~\eqref{eq:hierarchy-core} is therefore not to collapse these notions into one another, but to isolate precisely which implications are unconditional in the present framework and which require additional assumptions.

A first immediate consequence of this hierarchy is worth stating explicitly. Since epistemic completeness implies every notion that appears to its right in Eq.~\eqref{eq:hierarchy-core}, the contrapositive gives a direct route back: any operational phenomenon that rules out bounded ontological distinctness on preparations, or preparation noncontextuality, already rules out epistemic completeness. Under the additional standard assumptions discussed below, the same holds for the stronger conditional notions as well. In this precise sense, epistemic incompleteness provides a common umbrella under which several familiar departures from classical explanation can be understood.

\subsection{Epistemic completeness implies bounded ontological distinctness on preparations}

Bounded ontological distinctness for preparations requires that operational distinguishability be preserved as ontic distinctness \cite{Chaturvedi2020quantum}. In the present notation, this means preservation of the distinguishability task-values \(D_{n,1}\), and in particular of the two-state quantities \(D_{2,1}\) for every pair of preparations.

This implication is immediate from the definition of epistemic completeness, but it is conceptually central enough to record explicitly.

\begin{proposition}
\label{prop:ec-implies-bod}
Epistemic completeness implies bounded ontological distinctness on preparations.
\end{proposition}

\begin{proof}
By definition, epistemic completeness requires exact preservation of every empirical preparation-property of the form \eqref{eq:generic-task}. The distinguishability quantities \(D_{n,1}\) are a particular subfamily of such properties, corresponding to exact label identification tasks. Therefore epistemic completeness preserves \(D_{n,1}\) for every finite preparation set. In particular, it preserves \(D_{2,1}\) for every pair of preparations, which is exactly bounded ontological distinctness on preparations.
\end{proof}

The point is not merely that bounded ontological distinctness is contained in the present framework. More strongly, epistemic completeness is its natural extension from one distinguished operational quantity to the full family of empirical communication-task quantities associated with a finite preparation set.

\subsection{Preparation noncontextuality as a consequence}

Preparation noncontextuality says that operationally equivalent preparations must be represented by identical epistemic states \cite{PhysRevA.71.052108}. In the present framework, this follows already from bounded ontological distinctness, and hence from epistemic completeness.

Suppose two preparations \(P_1\) and \(P_2\) are operationally equivalent. Then no operational test can distinguish them, so
\begin{equation}
D_{2,1}^{(\cO)}(P_1,P_2)=\frac12.
\end{equation}
If bounded ontological distinctness holds for preparations, then
\begin{equation}
D_{2,1}^{(\Lambda)}(\mu_1,\mu_2)=\frac12.
\end{equation}
But
\begin{equation}
D_{2,1}^{(\Lambda)}(\mu_1,\mu_2)
=
\frac12\int_\Lambda d\lambda\,\max\{\mu_1(\lambda),\mu_2(\lambda)\}.
\end{equation}
Since \(\mu_1\) and \(\mu_2\) are normalized probability densities,
\begin{equation}
\int_\Lambda \max\{\mu_1,\mu_2\}\,d\lambda
=
1+\TV(\mu_1,\mu_2),
\end{equation}
where \(\TV\) denotes the total variation distance. Hence the value \(1/2\) is attained if and only if \(\TV(\mu_1,\mu_2)=0\), equivalently if and only if \(\mu_1=\mu_2\) almost everywhere. Strictly speaking, the conclusion is equality almost everywhere with respect to the reference measure on \(\Lambda\). Since epistemic states are identified up to null sets, we freely choose representatives and write this as \(\mu_0(\lambda)=\mu_1(\lambda)\) for all \(\lambda\).

Therefore:

\begin{corollary}
\label{cor:ec-implies-pnc}
Epistemic completeness implies preparation noncontextuality.
\end{corollary}

No specifically quantum assumption enters this implication. It is a direct consequence of exact preservation of distinguishability.

This consequence is important for the global interpretation of the framework. Any operational violation of preparation noncontextuality already implies violation of bounded ontological distinctness, and hence violation of epistemic completeness. Thus the present framework does not compete with preparation noncontextuality as an alternative classicality principle; it subsumes it as a consequence of a stronger preparation-side preservation requirement.

\subsection{Conditional links to maximally \(\psi\)-epistemic and Kochen--Specker notions}

For pairs of pure quantum states, bounded ontological distinctness is closely tied to the familiar overlap constraints associated with maximally \(\psi\)-epistemic structure \cite{PhysRevLett.110.120401,PhysRevLett.112.250403,Quanta22}. The present framework therefore inherits this connection on the two-state level. In particular, whenever one restricts attention to pure-state pairs in the standard ontological-model setting, epistemic completeness implies preservation of the corresponding distinguishability relations and therefore implies the associated maximal-overlap condition.

This is one reason the Kochen--Specker model is such a natural test case. It is the paradigmatic qubit model that comes as close as possible to maximally \(\psi\)-epistemic behavior while still revealing, through the equality witness, the hidden excess task-power required by ontological completion.

Once one moves from pairwise overlap structure to the broader hierarchy of contextuality notions, additional assumptions enter. In the usual ontological-model framework for pure states and projective measurements, preparation noncontextuality connects conditionally to Kochen--Specker noncontextuality through the familiar self-duality assumptions emphasized in the literature \cite{PhysRevLett.110.120401,Quanta22}. Thus, within that setting,
\begin{equation}
\text{epistemic completeness}
\Longrightarrow
\text{preparation noncontextuality}
\overset{\text{standard assumptions}}{\Longrightarrow}
\text{Kochen--Specker noncontextuality}.
\end{equation}
The same logic should be read carefully. The implication to preparation noncontextuality is unconditional in the present framework. The further implication to Kochen--Specker noncontextuality is conditional on the standard additional assumptions of that setting. Accordingly, whenever those assumptions are in force, any violation of Kochen--Specker noncontextuality also certifies epistemic incompleteness.

\subsection{Relation to Bell-type notions of classical explanation}

Bell local causality belongs to a different operational setting, but it is again a no-fine-tuning classicality principle, now attached to multipartite spacelike separated experiments. The present work does not claim that epistemic completeness and Bell local causality are equivalent, nor that one can simply identify their violations. The relation is instead structural and hierarchical: epistemic completeness is a preparation-side preservation principle that sits naturally above bounded ontological distinctness and preparation noncontextuality in the broader realist landscape of classical explanations.

This is exactly the sense in which the present framework is unifying. Several familiar notions of classicality single out one operational feature and ask whether it can be preserved in an ontological explanation. Epistemic completeness asks the same question at the level of the full family of empirical communication-task quantities of a preparation set. The resulting witness therefore organizes, under one preparation-side umbrella, the operational ingredients from which several earlier notions of classicality arise.

\subsection{Scope and strictness of the hierarchy}

It is useful to see how the hierarchy acts on concrete ontological models.

\paragraph*{Beltrametti--Bugajski model.}
In the Beltrametti--Bugajski model the ontic state is essentially the pure quantum state itself \cite{BBModel}. Distinct pure states therefore correspond to disjoint ontic supports. Consider the pair \(\ket{0}\) and \(\ket{+}\). Their operational distinguishability is
\begin{equation}
D_{2,1}^{(\cQ)}(\ket{0},\ket{+})
=
\frac12\left(1+\frac1{\sqrt2}\right),
\end{equation}
whereas their ontic distinguishability in the model is \(1\). Hence the model is epistemically incomplete already at the two-state level.

\paragraph*{Kochen--Specker model.}
The Kochen--Specker model is subtler. It reproduces pure-qubit projective statistics and is genuinely \(\psi\)-epistemic, yet the equality witness shows that even here ontological completion exposes additional hidden communication power. The point of Sec.~IX was that this hidden excess is not vague: for the trine and tetrahedral ensembles, the model realizes exactly the amount predicted by the equality gap.

These examples illustrate two qualitatively different modes of incompleteness. In \(\psi\)-complete models the hidden excess already appears at the level of pairwise distinguishability. In \(\psi\)-epistemic models it can instead be distributed across higher set-distinguishability tasks. The equality witness captures both possibilities within one framework.
\subsection{Relation to other no-fine-tuning frameworks}

The no-fine-tuning notion used here should not be conflated with the causal-faithfulness condition studied by Wood and Spekkens \cite{Wood_2015}. There the question is whether observed conditional independences admit a causal explanation without delicate parameter tuning in an underlying directed acyclic graph. Here the question is different: whether the empirically accessible communication-task properties of a finite preparation set can be preserved exactly at the ontic level without having to suppress part of the task-power available from the ontic state.

The closest structural relation is to the operational fine-tuning framework of Catani and Leifer \cite{catani2020mathematical}, which shows that operational no-fine-tuning conditions can be formulated already at the level of ontic extensions, prior to imposing the full causal structure of ontological models. The present framework is fully consistent with that viewpoint. What it adds is a specific and broad class of empirically accessible preparation-properties---one-way communication-task quantities---together with an exact equality witness extracted from a canonical family of them.

At the same time, modern prepare-and-measure work has developed several complementary task-based routes to nonclassicality: communication-game witnesses of preparation contextuality \cite{PhysRevLett.119.220402}, classicality certification directly from observable behaviors \cite{PRXQuantum.2.030311}, structural criteria for arbitrary prepare-and-measure scenarios \cite{GittonWoods2022}, and informationally restricted communication frameworks under natural physical assumptions \cite{VanHimbeeck2017,Tavakoli2020informationally,Tavakoli2022informationally,PauwelsPironioTavakoli2025}. These frameworks are not identical to the present one, but together they show that task-based and prepare-and-measure formulations of classicality now form a mature and interconnected line of research.

The same operational language also leads naturally to the complementary benchmark-constrained framework discussed earlier. Here the question is one of exact preservation under complete classical explanation. But the very same empirical communication-task quantities can also be used operationally, by bounding some tasks and witnessing nonclassicality through performance in others. In that broader framework, communication tasks play a dual role: they function simultaneously as witnesses and as constraints.

Taken together, these relations sharpen the unifying message of the paper. Epistemic completeness is not merely one more criterion beside existing notions. It extends distinguishability-based classicality from one selected operational quantity to the full family of empirical communication-task quantities associated with a finite preparation set. Consequently, whenever quantum theory departs from any classicality notion lying below it in the implication hierarchy, that departure is already evidence of epistemic incompleteness. In this way, the framework supplies a common language for several distinct failures of classical explanation within one exact operational umbrella.

\section{Conclusion}

We introduced epistemic completeness as a notion of classicality tailored to the following question: can the empirically accessible properties of a finite preparation set be reproduced exactly by an ontological or hidden-variable completion, or must every such completion inevitably contain additional hidden structure that is not operationally accessible?

Our answer is encoded by an exact witness. For the canonical family of set-distinguishability tasks, every epistemically complete theory satisfies
\begin{equation}
\avgD_n=\avgS_n
\end{equation}
for every finite preparation set. The proof is elementary at its core, but its consequence is strong: any nonzero deviation certifies epistemic incompleteness and quantitatively lower-bounds the amount of hidden ontic communication power that every ontological completion must conceal. The witness therefore does not add an isolated inequality to the literature; it expresses a preservation principle under which several familiar preparation-side notions of classical explanation appear as weaker descendants.

Quantum theory violates this equality in both directions. The trine and tetrahedral qubit ensembles furnish exact positive-direction violations; mixed-state higher-dimensional ensembles realize opposite-sign violations; and the trine and tetrahedral points are, respectively, the exact and numerically certified positive-direction maximizers of the \(n=3\) and \(n=4\) equalities. We further showed that these violations persist throughout explicit visibility and leakage families, down to arbitrarily low positive visibility and up to arbitrarily large leakage short of complete disclosure. Thus the operational quantum description is epistemically incomplete already on simple and highly symmetric qubit fragments, and that incompleteness remains robust deep into noisy and high-leakage regimes.

The Kochen--Specker \(\psi\)-epistemic model makes the witness quantitatively sharp. For the trine and tetrahedral ensembles, it exactly saturates the hidden ontic excess implied by the equality violation. In these canonical cases, the witness is therefore not merely a qualitative no-go statement against complete classical explanation; it identifies the exact amount of additional ontic communication power that ontological completion must expose.

The same nonzero deviation also has immediate operational and structural consequences. Because unrestricted classical communication models are epistemically complete, every nonzero deviation yields benchmark-constrained quantum communication advantage. Because every finite-dimensional commuting quantum theory is epistemically complete, a nonzero deviation is also a dimension-independent witness of coherence of the preparation set. And in any concrete realization of the certifying task values, it witnesses incompatibility of the certifying measurement family.

Finally, the framework clarifies the place of earlier notions of classicality. Epistemic completeness subsumes bounded ontological distinctness on preparations and thereby lies above preparation noncontextuality in the corresponding hierarchy. Under the standard additional assumptions familiar from the ontological-model literature, it also connects naturally to maximally \(\psi\)-epistemic structure and Kochen--Specker noncontextuality, while sitting alongside Bell local causality in the wider realist landscape of classical explanations. The resulting picture is unified and exact: the incompleteness of the quantum description is witnessed by the failure of a task-based equality that every complete classical explanation would have to satisfy. It would be especially interesting to develop semi-device-independent quantum key distribution under such operational communication-task constraints, and to determine how the resulting epistemic-incompleteness witnesses translate into robust security guarantees \cite{SDI1,SDI2,SDI3}.
\subsection*{Acknowledgments}
This project is funded by KLAR Grant NO BNI/PST/2023/1/00013/U/00001 funded by NAWA.
DS acknowledges support from STARS (STARS/STARS-2/2023-0809), Govt. of India.

\bibliographystyle{apsrev4-2}
\bibliography{FINAL_updated}

\end{document}